\documentclass[sn-mathphys]{sn-jnl_SA}

\usepackage{gensymb}	

\begin{document}

\title[Cellular design patterns]{Design Patterns of Biological Cells}

\author*[1]{\fnm{Steven S.} \sur{Andrews}}\email{steven.s.andrews@gmail.com}

\author[2]{\fnm{H. Steven} \sur{Wiley}}\email{steven.wiley@pnnl.gov}

\author[1]{\fnm{Herbert} M. \sur{Sauro}}\email{hsauro@uw.edu}

\affil*[1]{\orgdiv{Department of Bioengineering}, \orgname{University of Washington}, \orgaddress{\city{Seattle}, \state{WA}, \country{USA}}}

\affil[2]{\orgdiv{Environmental Molecular Sciences Laboratory}, \orgname{Pacific Northwest National Laboratory}, \orgaddress{\city{Richland}, \state{WA}, \country{USA}}}

\abstract{Design patterns are generalized solutions to frequently recurring problems. They were initially developed by architects and computer scientists to create a higher level of abstraction for their designs. Here, we extend these concepts to cell biology in order to lend a new perspective on the evolved designs of cells' underlying reaction networks. We present a catalog of 21 design patterns divided into three categories: creational patterns describe processes that build the cell, structural patterns describe the layouts of reaction networks, and behavioral patterns describe reaction network function. Applying this pattern language to the \textit{E. coli} central metabolic reaction network, the yeast pheromone response signaling network, and other examples lends new insights into these systems.}


\maketitle

\section{Introduction}

In 1966, the architect Christopher Alexander wrote ``The pattern of streets'' in which he proposed that urban street patterns should have bands of parallel one-way roads separated by perpendicular freeways~\cite{Alexander_1966}, showing that this would reduce driving times. No cities followed this advice, to our knowledge, but the paper was influential because it introduced the concept of creating novel patterns to solve specific problems. His subsequent work introduced a ``pattern language''~\cite{Alexander_1977, Alexander_1979}, with each of 253 patterns giving the principle of a solution but not the implementation. In Alexander's words, ``each pattern describes a problem which occurs over and over again in our environment, and then describes the core of the solution to that problem, in such a way that you can use this solution a million times over, without ever doing it the same way twice.'' \cite{Alexander_1977}.

These pattern concepts resonated in computer science, where Gamma et al. realized that programmers often solve the same problems repeatedly and could benefit from a set of common solutions. They described 23 software development ``design patterns'' that have been very influential in computer science~\cite{Gamma_Vlissides_1995}. These patterns offer a higher level of abstraction in software development than had been available previously, which helps individual programmers grasp the entirety of a project and also gives software engineers a common language for discussing their work.

As with computational systems, biological organisms also create high-level complex functions from simple processes, are hierarchically organized, and perform sophisticated information processing~\cite{McAdams_Shapiro_1995, Hartwell_Murray_1999, Lazebnik_2002, Bhalla_2003, Sauro_Kholodenko_2004, Bray_2009}. Additionally, organisms are similar to human-engineered systems in that they are confronted by similar problems repeatedly, which has often led to similar solutions (i.e. convergent evolution). Based on these parallels, we explored whether a design pattern concept could be meaningfully applied to cell biology.

Our work builds on substantial prior work that has investigated recurring biological structures. Alon and coworkers investigated statistically over-represented structures, which they call network motifs \cite{Milo_Alon_2002, Milo_Alon_2004}. They found, for example, that a three-component feedforward loop occurs significantly more often in real biological networks than in random ones. Additionally, several researchers have classified common network structures by function, showing how particular mechanisms are able to create particular functions \cite{Tyson_Novak_2003, Sauro_Kholodenko_2004}. Also, substantial work has focused on investigating the modular nature of natural biochemical networks, where a module is a subsystem that behaves sufficiently independently of other system portions that it retains its intrinsic properties irrespective of what it is connected to \cite{Hartwell_Murray_1999, DelVecchio_Sontag_2008, Sauro_2008}. That work has generally supported a modular view of cell biology, although with the recognition that there is invariably some retroactivity and other crosstalk. Design patterns are closely related to motifs, mechanisms, and modules, but have the distinction that they are explicitly solutions to problems. Nevetheless, these concepts are not exclusive, so design patterns can also be motifs, mechanisms, or modules.

Here, we present a catalog of cell biology design patterns, which Table~\ref{table:PatternList} summarizes. Because design patterns exist at all levels of detail within biology, from the designs of specific molecules to the designs of multi-cellular organisms, we restrict this work to the chemical reaction networks that animate individual cells. Further, we only consider patterns that represent solutions to specific biological problems, and that are represented across multiple organisms or cell types. The resulting catalog is divided into the same categories as are used for computer science~\cite{Gamma_Vlissides_1995}: (1) creational patterns, which provide object creation mechanisms, (2) structural patterns, which represent interconnections among the objects, and (3) behavioral patterns, which describe the objects' dynamical behaviors. Our catalog is necessarily subjective, reflecting the experiences and biases of the authors, so we encourage others to modify this list as they see fit, or to create their own.

\begin{table}[h]
\begin{center}
\caption{List of cellular design patterns}
\label{table:PatternList}
\begin{tabular}{@{}lll@{}}
\toprule
Creational & Structural  & Behavioral\\
\midrule
Template & Input/output & Adaptation\\
Assembly line & Collector/broadcaster & Periodic\\
Passive assembly & Common currency & Proportional output\\
Active assembly & Chain & Hyperbolic output\\
Pores and pumps & Parallel paths & Switching\\
Transformation  & One-way cycle & Direction maker\\
  & Annotation & Insulator\\
  & & Fold-change/ratiometric\\
\botrule
\end{tabular}
\end{center}
\end{table}

\section{Catalog of design patterns}

\subsection{Creational patterns}

\begin{figure}[ht]
\centering
\includegraphics[height=4cm]{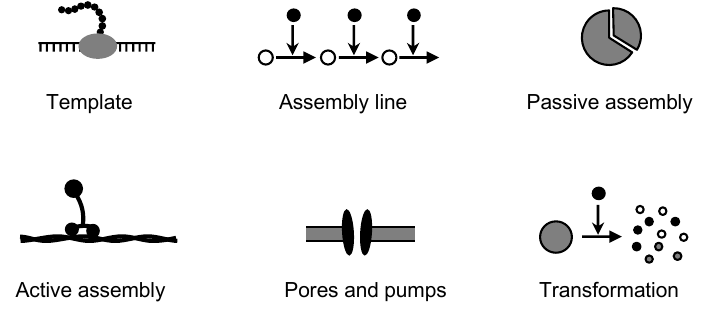}
\caption{Cartoons of creational patterns, each of which represents an iconic example of the given pattern.}
\label{fig:CreationalPatterns}
\end{figure}

We define creational patterns as the solutions that cells use to create the physical objects that they are built from (Figure \ref{fig:CreationalPatterns}). The primary objects in a cell are nucleic acids, proteins, lipids, and small molecules, and those are then assembled into macromolecular structures such as membranes, cytoskeletal filaments, protein complexes, and lipid rafts.

\subsubsection{Template}

\textit{Problem.} Cells need a diverse set of macromolecules (i.e. DNA, RNA, and proteins) that are built from pre-specified designs, and are heritable and evolvable.

\textit{Solution.} Biosynthesis using a master copy of the macromolecule sequence, which is then faithfully copied using a relatively small set of enzymes. Kinetic proofreading steps~\cite{hopfield1974kinetic}, which consume energy, are necessary for improving copying fidelity over the best that could be achieved in a copying system that does not consume energy.

There are three dominant versions of this pattern, which are DNA replication, DNA transcription to RNA, and RNA translation to proteins. Each is performed by complex biochemical machinery that moves along the template and catalyzes the production of the newly synthesized molecule, and each includes its own version of kinetic proofreading~\cite{Boeger_2022}. The fact that this template pattern is mutable, is heritable, and impacts phenotype, appears to be essential to the evolution of complex life. Nevertheless, this pattern is not essential for the mere short-term survival of a cell, as evidenced by the survival of cells that don't contain DNA, including red blood cells, lens fiber cells, and platelets.

\subsubsection{Assembly line}

\textit{Problem.} Cells require molecules that perform specific physical or chemical functions that are beyond the capabilities of nucleic acids or proteins. These molecules include lipids, polysaccharides, polyamines, protein cofactors, metabolites, and many small molecules.

\textit{Solution.} Biosynthesis uses an assembly line of enzymes, each of which performs a specific chemical reaction. These assembly lines can have incoming branches, outgoing branches, or cycles, as needed for managing chemical fluxes.

\begin{figure}[ht]
\centering
\includegraphics[height=8cm]{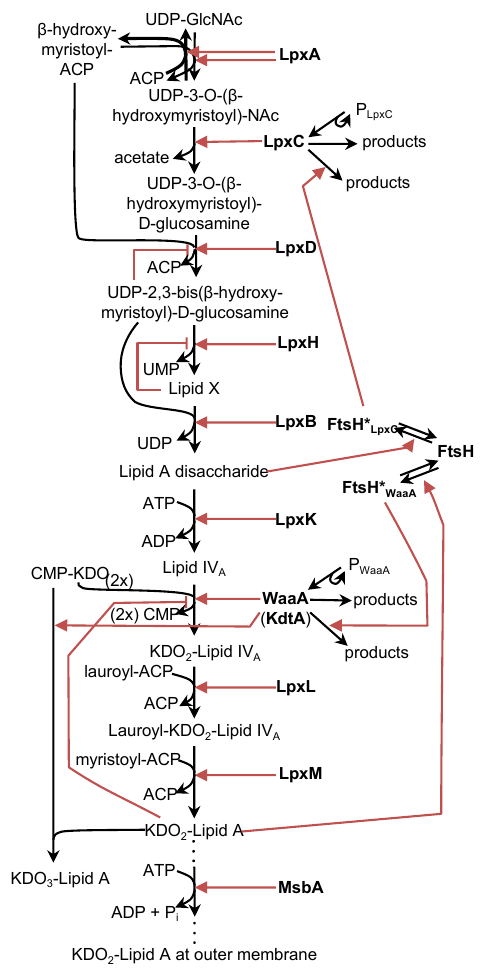}
\caption{Example of an Assembly Line pattern. This is the \textit{E. coli} biosynthesis pathway for lipopolysaccharide. Black arrows represent chemical reactions in which reactants are converted to products and red arrows represent enzymatic or other influences on reaction rates. Figure modified with permission from ref. \cite{Emiola_Andrews_2015}.}
\label{fig:LPSpathway}
\end{figure}

This pattern describes both anabolic and catabolic metabolic pathways. A notable aspect of such assembly lines is that they require negative feedback loops within them in order to match the supply created by the assembly line to the demand for its products~\cite{Hofmeyr_Cornish-Bowden_2000}. Figure \ref{fig:LPSpathway} shows a typical example of this pattern, in this case illustrating the \textit{E. coli} biosynthesis of lipopolysaccharide \cite{Emiola_Andrews_2015}, which is the dominant outer membrane lipid in these bacteria. Note the prevalence of negative feedbacks, which are thought to be used to match supply with demand~\cite{curien2009understanding}. 

\subsubsection{Passive assembly}

\textit{Problem.} The molecules and macromolecules that compose a cell typically require assembly into transient or permanent structures.

\textit{Solution.} Thermodynamically favorable self-assembly in which the separate components diffuse randomly until they encounter each other and then bind together spontaneously. The lack of energy input tends to make these processes reversible.

Examples of intracellular passive assembly arise at many size scales. They include the spontaneous folding of some proteins, such as the widely studied RNase A \cite{Anfinsen_1973}, many protein dimerization and oligomerization reactions \cite{Marianayagam_Matthews_2004}, the formation of amyloid fibrils, and protein self-assembly into cytoskeletal filaments \cite{Rivas_Minton_2001, Gardner_Odde_2011}. Extended protein structures can also assemble through passive assembly, including the well-organized \textit{E. coli} and EGF receptor clusters~\cite{Goldman_Bray_2004, Mayer_Yu_2018} and the spontaneous phase separations that create the more loosely organized cytoplasmic droplets~\cite{Brangwynne_Hyman_2009} and lipid rafts \cite{Simons_Sampaio_2011}. Additionally, transcription factor binding to target DNA sequences occurs through passive assembly~\cite{Schmidt_Lipkow_2014}.


\subsubsection{Active assembly}

\textit{Problem.} Transient or permanent macromolecular structures do not always self-assemble effectively, due to either unfavorable thermodynamics or excessively slow kinetics.

\textit{Solution.} Assembly with energy input and, typically, assistance from other cell components. These processes are often effectively irreversible.

Any macromolecular assembly or localization process that requires ATP hydrolysis, GTP hydrolysis, or some other energy input falls in the category of active assembly. As with passive assembly, active assembly also occurs at many size scales. Examples include protein folding that is catalyzed by foldases or other chaperones~\cite{Sorokina_Koonin_2022}, cofactor insertion into proteins such as the heme group into nitric oxide synthase, myoglobin, or hemoglobin \cite{Ghosh_Stuehr_2011, Tupta_Ghosh_2022}, intracellular trafficking using motor proteins that move along actin or microtubules~\cite{Howard_2001}, control of membrane curvature using integral or peripheral membrane proteins~\cite{McMahon_Gallop_2005}, and growth of actin filaments to drive cell motility~\cite{Pollard_Mullins_2000}.

\subsubsection{Pores and pumps}

\textit{Problem.} Cellular components, from ions to proteins, typically need to be localized to the correct sides of membranes, including the plasma membrane, nuclear membrane, and other organelle membranes.

\textit{Solution.} Trans-membrane pores and pumps that use either active or passive transport. These pores and pumps are typically quite selective about what molecules they transmit and are often gated by external signals.

Cell membranes are quite permeable to oxygen, carbon dioxide, and other small nonpolar molecules but are effectively impermeable to larger and more charged species, a property that is essential to establishing and maintaining cell organization. Transport of these latter species occurs via transporters and channels~\cite{Dubyak_2004}, including ion channels, passive and active transporters for ions or other small molecules, proton pumps, ABC transporters, photosynthetic reaction centers for electron transport, and ATP synthase proteins for mitochondrial proton transport. The nuclear pore complex is a particularly large pore, which enables passive transport of small molecules and performs active transport on proteins that carry nuclear localization or nuclear export signals \cite{Wente_Rout_2010}.

\subsubsection{Transformation}

\textit{Problem.} All cellular components need to be either disassembled or transformed into new components at some point.

\textit{Solution.} Proteins that degrade or transform cellular components. These proteins require tight regulation to ensure that they only degrade or transform the correct components.

Cellular component degradation occurs in both the cytoplasm and lysosomes. In the former case, proteins are typically tagged for degradation with ubiquitin groups, followed by degradation in proteasomes, which are large protease complexes \cite{Hochstrasser_1995, Gu_Greenberg_2023}. In the latter case, cellular components are transported to lysosomes and then degraded by hydrolytic enzymes including proteases, lipases, RNases, DNases, glycosidases, and phosphatases. The rest of the cell is protected from these enzymes both because they are sequestered to lysosomes, and because they require the highly acidic lysosome environment to function. All degradation is tightly regulated, including protein degradation during autophagy \cite{Klionsky_Emr_2000}, DNA degradation during development and apoptosis \cite{Nagata_2005}, and lipid degradation to preserve energy homeostasis \cite{Liu_Czaja_2013}.

Some proteases can also transform particular proteins into other functional proteins. For example, zymogens are inactive enzyme precursors that are subsequently converted to their active forms through selective enzymatic cleavage~\cite{Neurath_Walsh_1976}. Transformation of yet other proteins, such as neuropeptides or membrane-anchored growth factors~\cite{rouille1995proteolytic, singh2005autocrine}, provides an efficient regulatory mechanism that both arrests the action of one protein and replaces it with the function of a modified one.

\subsection{Structural patterns}

\begin{figure}[ht]
\centering
\includegraphics[height=3.5cm]{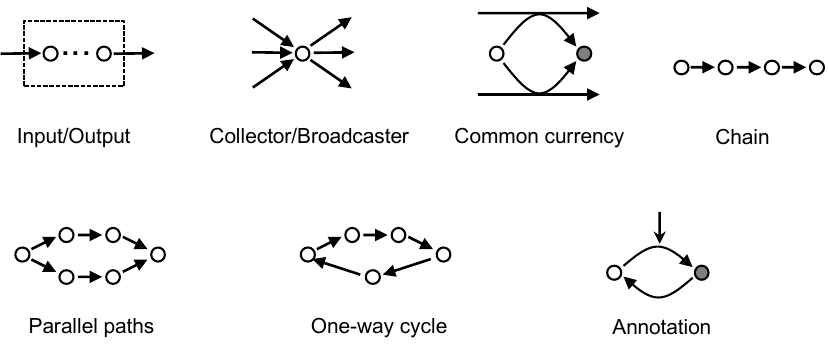}
\caption{Cartoons of structural patterns, with each one representing an iconic example of the given pattern.}
\label{fig:StructuralPatterns}
\end{figure}

\begin{figure}[ht]
\centering
\includegraphics[height=7cm]{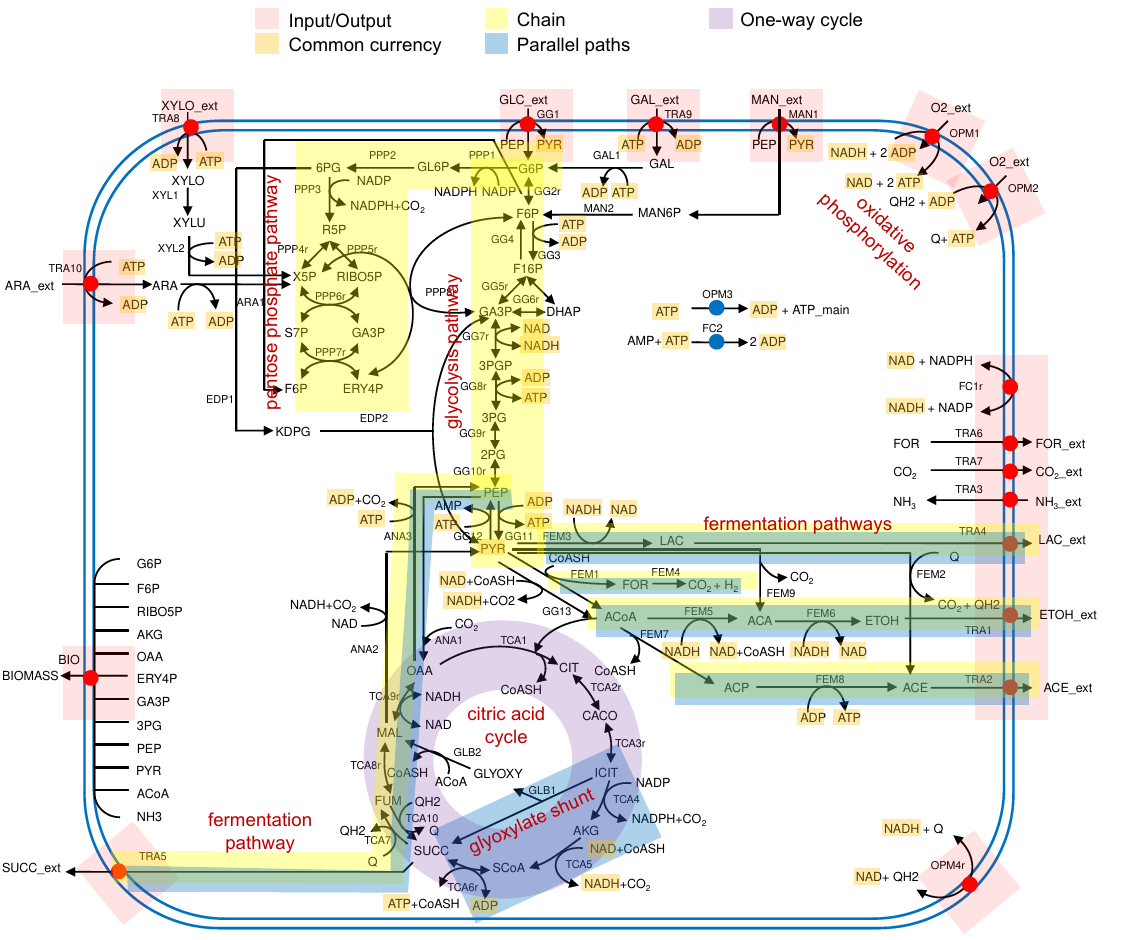}
\caption{The central metabolic system for E. coli with highlighted regions showing several structural patterns. The metabolic map is reprinted with permission from ref. \cite{Trinh_Srienc_2008}.}
\label{fig:MetabolicMap}
\end{figure}

\begin{figure}[ht]
\centering
\includegraphics[height=7cm]{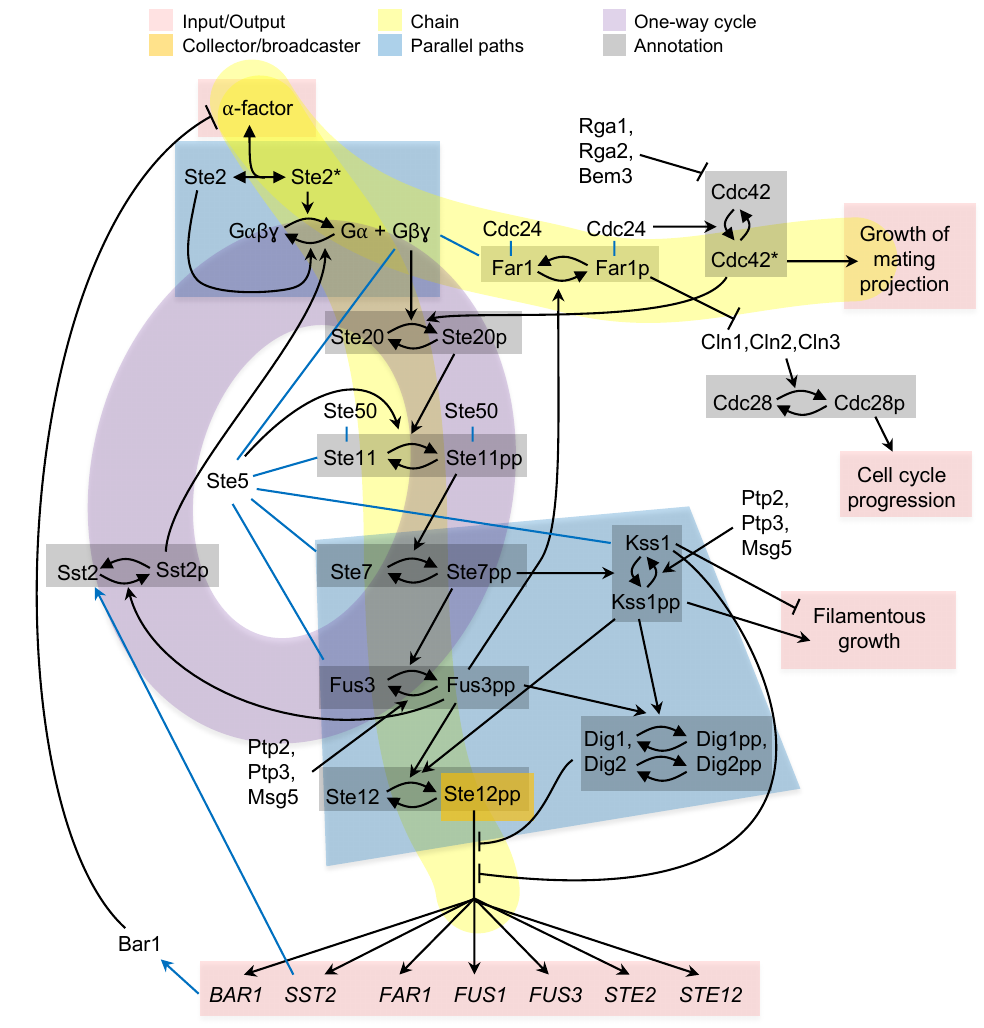}
\caption{Diagram of the \textit{Saccharomyces cerivisiae} pheromone response signaling pathway. This diagram was created from information described in refs. \cite{Kurjan_1992, Wang_Dohlman_2004, Bardwell_2005, Andrews_Arkin_2007, Bush_Colman-Lerner_2016}.}
\label{fig:SignalingMap}
\end{figure}

We define structural design patterns as solutions that are addressed by the topological structure of the cell's biochemical reaction networks (Figure \ref{fig:StructuralPatterns}). We focus on metabolic and signaling networks, although genetic regulatory~\cite{Bolouri_Davidson_2002, Karlebach_Shamir_2008} and other intracellular networks may use the same design patterns as well. Figures~\ref{fig:MetabolicMap} and~\ref{fig:SignalingMap} put these patterns into context by highlighting structural patterns in the \textit{E. coli} central metabolic system and the yeast pheromone response signaling system.

\subsubsection{Input/output}


\textit{Problem.} Chemical reaction networks in living organisms are not closed systems but must interface with the outside world.

\textit{Solution.} Specific elements that input material and/or information from the environment. If they input material, then they also need to output material to maintain mass balance; if information, then they also need to output information for the system to serve a purpose.

Metabolic inputs and outputs occur primarily through passive diffusion and membrane transporters (Figure~\ref{fig:MetabolicMap}, pink highlighting). Signaling inputs often occur through the detection of extracellular ligands by specific receptors, while the outputs comprise specific cell actions, such as growth in a specific direction or arrest of the cell cycle (Figure~\ref{fig:SignalingMap}, pink highlighting). Signaling inputs can also be internal to a cell; for example, heat shock proteins and systems that sense cell nutrient levels, osmotic conditions, and pH all input information within a cell.

\subsubsection{Collector/broadcaster}

\textit{Problem.} Many separate cell functions need to be regulated simultaneously in a consistent manner.

\textit{Solution.} Information flow periodically converges at central nodes that then provide consistent information to multiple downstream targets.

This pattern, which applies exclusively to information flow, represents what is often termed a master regulator~\cite{Chan_Kyba_2013}, meaning a protein or gene that regulates many downstream targets. Examples include many transcription factors (Figure \ref{fig:SignalingMap}, orange highlighting), sheddases~\cite{lichtenthaler2018proteolytic}, TOR and mTOR proteins~\cite{Loewith_Hall_2011}, and hormones; all of these collect information from multiple inputs and broadcast it to many targets. Allostery and conformational spread~\cite{Duke_Bray_2001} interactions are mechanistically different from these examples but also represent collector/broadcaster patterns because they also combine information from multiple inputs. These signaling control points often have a relatively low abundance compared with their interaction partners, making them limiting in regulatory pathways. Additionally, many are highly phosphorylated, which is thought to control their activity~\cite{shi2016conservation}.



\subsubsection{Common currency}

\textit{Problem.} Many cell processes have similar energetic or chemical requirements.

\textit{Solution.} Develop a few standard energy and chemical sources and use them repeatedly.

Common currency examples include ATP and GTP, which act as energy sources, NAD$^+$/NADH, which transport electrons, and pyruvate, which is a common metabolic precursor (Figure~\ref{fig:MetabolicMap}, orange highlighting). Additionally, the phosphate group is widely used as a common currency for information transmission in signaling networks (Figure~\ref{fig:SignalingMap}, not highlighted).

This pattern is topologically identical to the collector/broadcaster pattern but is listed separately because it solves a different problem. This pattern is also essentially the same as the bow-tie architecture that Csete and Doyle identified~\cite{Csete_Doyle_2004}, where their name represents the fact that many nutrients are converted to the single common currency at the ``knot'' of the bow-tie, which is then applied to many uses. They pointed out that the use of a common currency has the benefits of facilitating tight regulation, damping out transient fluctuations, and supporting highly specialized and efficient metabolism. It also supports evolvability of upstream and downstream components by improving modularity.

\subsubsection{Chain}

\textit{Problem.} Many signals and metabolites need to be processed through a defined sequence of steps.

\textit{Solution.} A linear chain of chemical reactions, possibly with incoming or outgoing branches.

Many biochemical reaction networks are roughly linear, with an input at one end, an output at the other end, and a chain of steps in between. Metabolic examples include the glycolysis pathway, pentose phosphate pathway, and several fermentation pathways (Figure~\ref{fig:MetabolicMap}, yellow highlighting). Signaling examples often extend from ligand detection at one end to transcriptional regulation at the other end, but can also connect other inputs and outputs (Figure~\ref{fig:SignalingMap}, yellow highlighting). This Chain pattern is closely related to the Assembly line creational pattern, but focuses on the network structure rather than the specifics of how the end product is manufactured.

\subsubsection{Parallel paths}

\textit{Problem.} The Chain pattern can be too limiting if it includes steps that are infeasible or have inadequate sensitivity in particular situations.

\textit{Solution.} Multiple parallel paths that complement each other.

Metabolic parallel paths typically address variations in nutrient availability. For example, aerobic organisms produce ATP with oxidative phosphorylation when they have sufficient oxygen and otherwise use less efficient fermentation pathways that produce succinate, lactate, formate, ethanol, and/or acetate, depending on nutrient availability (Figure~\ref{fig:MetabolicMap}, blue highlighting). Similarly, the glyoxylate cycle is a bypass in the citric acid cycle, which functions if some simple sugars are unavailable.

In signaling networks, multiple parallel paths enable more versatile information processing than any single pathway could accomplish. These parallel paths include feedforward loops, which have been shown to perform complex behaviors such as fold-change detection~\cite{Goentoro_Alon_2009} and sign-sensitive delay~\cite{Mangan_Alon_2003}. In yeast signaling, ligand-bound Ste2 promotes G-protein dissociation and, in parallel, ligand-unbound Ste2 promotes G-protein association~\cite{Bush_Colman-Lerner_2016} (Figure~\ref{fig:SignalingMap}, blue highlighting). Together, these two paths enable linear signal transmission~\cite{Andrews_Brent_2016} and cause G-protein activity to reflect the fraction of occupied receptors rather than the number of them \cite{Bush_Colman-Lerner_2016}.



\subsubsection{One-way cycle}

\textit{Problem.} Upstream portions of a network sometimes require material or information that is produced downstream.

\textit{Solution.} A one-way cycle, in which some of the downstream material or information is returned to an upstream location in the same pathway.

Metabolic cycles include the citric acid, urea, photorespiration, and Calvin cycles, all of which produce specific metabolites at one part of the cycle that are then consumed elsewhere in the same cycle. For example, Figure \ref{fig:MetabolicMap} (purple highlighting) shows that the citric acid cycle starts with a citrate (CIT) molecule, which undergoes several reactions to eventually form an oxoacetate molecule (OAA); that combines with an acetyl-CoA (ACoA) molecule to synthesize another citrate molecule and start a new cycle.

Signaling cycles can typically be characterized as either positive or negative feedback loops. Positive feedback generally creates amplification and, if strong enough, bistability. Negative feedback typically improves monostability and hence is essential for adaptation and homeostasis, although it can also create oscillations if it is coupled with some element that creates a delay~\cite{Sauro_2017}. The feedback loop highlighted in Figure \ref{fig:SignalingMap} (purple highlighting) represents a negative feedback that resets the signaling system after exposure to pheromone \cite{Dohlman_Thorner_1996, Venkatapurapu_Elston_2015}, contributes to ratiometric pheromone sensing \cite{Bush_Colman-Lerner_2016}, and reduces the impacts of stochasticity on the signaling pathway \cite{Dixit_Dohlman_2014}. Positive and negative feedback loops are discussed in more detail below in the Behavioral patterns section.

\subsubsection{Annotation}

\textit{Problem.} Macromolecules need to be in different states at different times, whether to represent information or to respond to varying cellular conditions.

\textit{Solution.} Reversible chemical modification that is catalyzed by enzymes that either carry information themselves or are responsive to current conditions.

This pattern represents both covalent and non-covalent modifications. The former include protein phosphorylation (Figure \ref{fig:SignalingMap}, gray highlighting), protein methylation~\cite{Sourjik_Berg_2002, Paik_Kim_2007}, DNA methylation \cite{Moore_Fan_2013}, and ribosome modification \cite{Sloan_Bohnsack_2017}. The latter includes nucleotide substitution and hydrolysis, of which examples include GTP substitution in G-proteins and GTPases (e.g. Cdc42 in Figure~\ref{fig:SignalingMap}), ATP substitution in the \textit{E. coli} MinD protein \cite{Lutkenhaus_2007}, and ATP substitution in actin filaments~\cite{Pollard_Mullins_2000}. These modifications are often thought about as simple tags that label protein or DNA states, but also clearly need to create chemical effects to be functional. For example, protein phosphorylation can change the protein conformation through electrostatic interactions, which then exposes an active site and hence alters functionality \cite{Groban_Jacobson_2006}.

Annotation can be regulated through either one-sided or two-sided control, meaning whether one or both reaction directions are externally regulated. As a typical example of one-sided control, protein phosphorylation is often regulated by the activity of an upstream kinase, while the corresponding phosphatase is constitutively active (e.g. refs.~\cite{Huang_Ferrell_1996, Sauro_Kholodenko_2004} and Ste20, Ste11, and Ste7 in Figure~\ref{fig:SignalingMap}). Protein phosphorylation can also have two-sided control, in which both the kinase and the phosphatase are regulated (e.g. the Fus3, Kss1, and Ste12 proteins in Figure~\ref{fig:SignalingMap}).

An open question concerns the role of highly phosphorylated proteins, which often have a dozen or more separate phosphorylation sites~\cite{Roach_1991, Mann_Pandey_2002}. Possibilities for their function include that they promote bistability~\cite{Kapuy_Novak_2009}, stabilize proteins that have many positively charged residues~\cite{McPherson_Turnbough_2010}, serve as a tagging mechanism for protein degradation~\cite{Rosen_Ron_2004}, or promote signaling specificity to multiple downstream signaling cascades~\cite{vanderGeer_Pawson_1996}.

\subsection{Behavioral patterns}

\begin{figure}[ht]
\centering
\includegraphics[height=4cm]{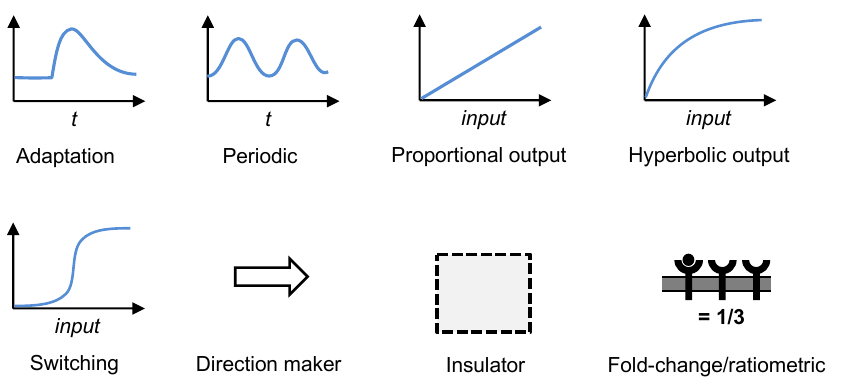}
\caption{Cartoons of behavioral patterns, with each one representing an iconic example of the given pattern.}
\label{fig:BehavioralPatterns}
\end{figure}

Behavioral patterns focus on the dynamics of biochemical reaction networks, including their input-output properties, temporal changes, adaptation, and stochastic amplification or reduction (Figure~\ref{fig:BehavioralPatterns}). They represent \textit{what} a cell does. As a result, all behavioral patterns depend at least to some extent on the rate constants of specific reactions.

Reaction networks exhibit an enormous range of behaviors, so this catalog only lists a selection of behavioral patterns that are very widely observed. It also focuses on patterns that can be modeled with non-spatial deterministic methods, thereby omitting patterns that relate to intracellular spatial organization~\cite{Moraru_Loew_2005, Dehmelt_Bastiaens_2010} and stochastic influences~\cite{Rao_Arkin_2002, Elowitz_Swain_2002, Tsimring_2014}.

\subsubsection{Adaptation}

\textit{Problem.} Many biochemical concentrations need to be maintained at essentially constant levels, and need to return to those levels after being disturbed.

\textit{Solution.} Biochemical networks that perform adaptation.

This pattern describes homeostatic systems, such as regulation of yeast cell volume despite osmotic shock perturbations~\cite{Muzzey_vanOudenaarden_2009}, mammalian glucose homeostasis despite metabolic fluctuations~\cite{Saunders_Wessels_1998}, and nitrate homeostasis in plants despite variations in soil nitrate levels~\cite{Huang_Ruoff_2012}. This pattern also describes situations where disturbances contain useful information and do perturb the system, but the system then adapts back to its baseline behavior. This latter case is exemplified by \textit{E. coli} chemotaxis, in which swimming cells change their tumbling likelihood upon exposure to attractant or repellant, but then adapt back to their baseline probability a few seconds later~\cite{Sourjik_Berg_2002}. Likewise, exposing yeast cells to $\alpha$-factor pheromone leads to pheromone responses but also causes the cell to secrete the Bar1 protease; the protease degrades extracellular pheromone, leading to attenuated signaling and hence adaptation \cite{Jin_Elston_2011}.

\begin{figure}[ht]
\centering
\includegraphics[height=3cm]{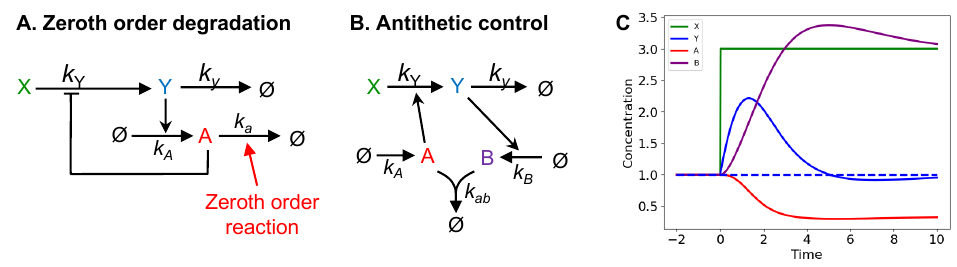}
\caption{Adaptation design pattern, showing two mechanisms for robust perfect adaptation. In each, X is the input and Y exhibits perfect adaptation to variation in X. (A) Zeroth order mechanism, in which the integral is stored in A. (B) Antithetic mechanism, in which the integral is stored in the concentration difference $B-A$. (C) Simulation of the antithetic mechanism. Here, all species start at concentration 1, $X$ increases to 3 at $t=0$, and $Y$ is perturbed but then adapts. Details: $k_A$ is zeroth order, $k_y$ is first order, and $k_Y$, $k_B$, and $k_{ab}$ are second order; $k_Y = k_y = k_A = k_a = k_{ab} = 1$. }
\label{fig:Adaptation}
\end{figure}

Adaptation invariably occurs through negative feedback and, conversely, negative feedback almost always produces some adaptation (an exception is when it produces oscillations, described below). However, perfect adaptation, meaning that the system adapts exactly back to a setpoint value even with a constant perturbation, is more difficult to achieve. It requires that the feedback loop adjusts the output based on the time integral of the difference between the system output and some setpoint value, which is called integral feedback control~\cite{Yi_Doyle_2000}. Figure~\ref{fig:Adaptation} shows the two mechanisms that have been found for robust perfect adaptation, where ``robust'' means that the perfect adaptation arises from the network structure rather than from finely tuned parameters. In both cases, the controlled system is shown at the top of the network, from X to Y to $\emptyset$, and the controller is below. In the first mechanism, the chemical species containing the integral information, A, decays through a reaction whose rate is independent of the chemical's concentration, which is called a zeroth order reaction~\cite{Yi_Doyle_2000, Qian_DelVecchio_2018}. In the second mechanism, called antithetic control, the difference is computed by comparing the production rates of two species by having them bind to each other in a 1:1 ratio~\cite{Briat_Khammash_2016}. Panel C shows perfect adaptation to a step input using the antithetic control network.

\subsubsection{Periodic}

\textit{Problem.} Many cell systems, especially those regulating cell division, need to operate in a periodic fashion.

\textit{Solution.} Biochemical reaction networks that can produce regular oscillations.

Biochemical oscillators have been observed in metabolism~\cite{Tu_McKnight_2006}, signaling pathways~\cite{Nelson_White_2004, Shankaran_Wiley_2009, Isomura_Kageyama_2014}, and genetic regulatory networks \cite{Leloup_Goldbeter_2003}, and are essential for such biological processes as cell division and circadian rhythms. Oscillators have also been engineered, of which the repressilator is the best known~\cite{Elowitz_Leibler_2000}. 

\begin{figure}[ht]
\centering
\includegraphics[height=5cm]{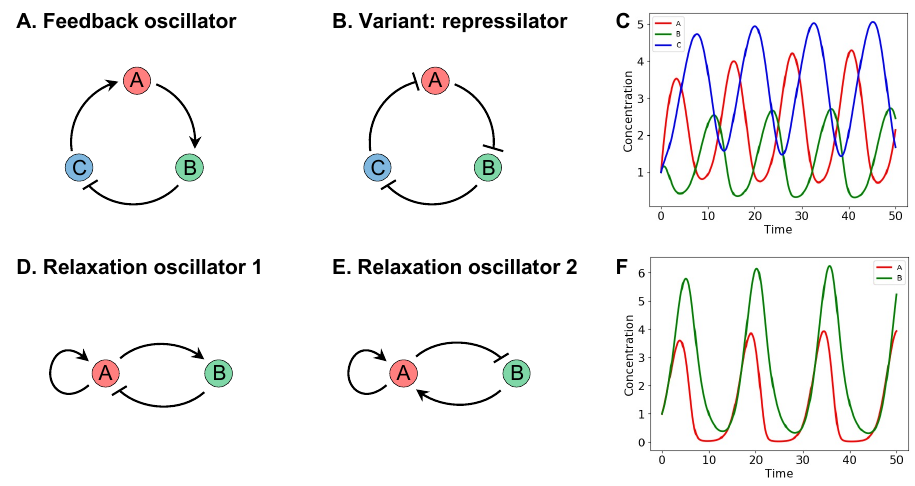}
\caption{Periodic design pattern, showing feedback oscillators in the top row and relaxation oscillators in the bottom row. Graphs in panels C and F show the repressilator and relaxation oscillator 1, respectively. Details: each node was represented by production and destruction reactions, e.g. $\emptyset \xrightarrow{k_A} A \xrightarrow{k_a} \emptyset$, with activations promoting the production reactions using first order kinetics and repressions promoting the destruction reactions with Michaelis-Menten kinetics (e.g. $K_A$ is the Michaelis constant for repression at node A). All species started with concentration 1. For panel C, $k_A=2$ and $k_B=k_C=k_a=k_b=k_c=K_A=K_B=K_C=1$. For panel F, $k_A = 1.1$, $k_a=k_B=1$, and $k_b=0.5$.}
\label{fig:Oscillator}
\end{figure}

There are two general patterns for creating oscillations, both of which require negative feedback at some point~\cite{Chickarmane_Sauro_2007, Novak_Tyson_2008}. The feedback or phase-shift oscillator, shown in Figure~\ref{fig:Oscillator}A-C, requires at least three states arranged in a cyclic structure and produces nearly sinusoidal oscillations. This system needs to have an odd number of negative influences over one cycle in order to create a net sign change in the signal; for example, panel A has one negative influence, and panel B has three negative influences. In the simplest version, there are equal time delays between each state's activation, each of which can have a 60\degree\ phase shift; combining this with the 180\degree\ phase shift from each negative influence produces a total phase shift of 360\degree\ (or some larger multiple of 360\degree) over the complete cycle. The oscillations are sustained so long as the gain around the cycle is greater or equal to one. Feedback oscillators are observed in NF-$\kappa$B and ERK signaling pathways~\cite{Nelson_White_2004, Shankaran_Wiley_2009}.

The relaxation oscillator, shown in Figure~\ref{fig:Oscillator}D-F, operates via a charging component, a negative feedback loop, and a bistable switch. The charging component gains charge in one switch state, triggers the switch to change states when it reaches a given threshold, and then gets rapidly discharged in this second switch state; afterward, the switch returns to its original state and the sequence repeats. In panels D and E, species A is the switch and species B is the charging component. Relaxation oscillators are found in the eukaryotic cell cycle \cite{Tyson_1991}, trains of intracellular calcium spikes that are elicited by hormonal or neurotransmitter signals \cite{Goldbeter_Berridge_1990}, and cAMP oscillations in \textit{Dictyostelium discoideum} cells \cite{Halloy_Goldbeter_1998}. In addition, the \textit{E. coli} Min system exhibits an intriguing spatiotemporal oscillation \cite{Lutkenhaus_2007}, which can also be described as a relaxation oscillator.

Although less well understood, many cells also exhibit regular pulsing in diverse systems that are not obviously tied to rhythmic behaviors. For example, the \textit{Bacillus subtilis} soil bacterium exhibits pulses in its genetic competence, sporulation initiation, and stress response pathways~\cite{Levine_Elowitz_2013}.

\subsubsection{Proportional output}

\textit{Problem.} Signal transmission within or between cells needs to occur with minimal information loss.

\textit{Solution.} Linear transmission, in which the output is directly proportional to the input, reduces information loss~\cite{sauro2007mapk,Andrews_Balazsi_2018}.

Linear signal transmission is widely observed in cell signaling systems, including in the yeast pheromone response system~\cite{Yi_Simon_2003, Yu_Brent_2008} and the mammalian EGF, Wnt, and Tgf$\beta$ signaling pathways~\cite{Nunns_Goentoro_2018}. Despite its conceptual simplicity and widespread occurrence, it is not trivial to produce \cite{Black_Leff_1983, Andrews_Brent_2016}

\begin{figure}[ht]
\centering
\includegraphics[height=5cm]{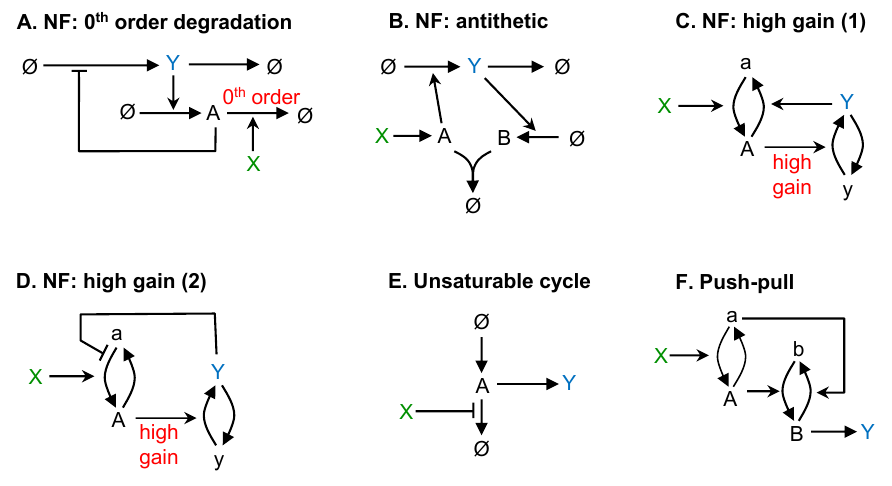}
\caption{Mechanisms for the Proportional output design pattern. In all cases, X is the input and the steady-state concentration of Y increases in direct proportion to X. ``NF'' stands for negative feedback.}
\label{fig:ProportionalOutput}
\end{figure}

Figure~\ref{fig:ProportionalOutput} shows several approaches for creating linear signal transmission with biochemical reactions~\cite{Andrews_Balazsi_2018}. In each case, X is the input and the steady-state concentration of Y can be shown to increase in direct proportion to X over some input range. Panels A and B show approaches in which negative feedback creates linearity in the same ways that it creates homeostasis in the Adaptation patterns. These include the use of a zeroth order degradation reaction or antithetic control, but the parameter that created the initial system's setpoint is replaced with a new system input. The antithetic approach was synthesized in a gene expression system that was confirmed to exhibit linearity \cite{Nevozhay_Balazsi_2009}. Panels C and D show approaches in which negative feedback creates linearity in a system that has a ``comparator-adjustor,'' which compares the input and output values, and then amplifies the difference to generate the output. In both panels, the cycle with ``a'' and ``A'' compares the output with the input and then adjusts the output as needed. The EGF signaling system appears to use this approach~\cite{sauro2007mapk,Sauro_2017,Nunns_Goentoro_2018}. This is also the most common means for producing linearity in analog electronic amplifiers, where such devices are called feedback amplifiers.

Panel E shows a particularly simple approach, which does not require feedback, in which the input modulates an unsaturated or unsaturable cycle through linear reaction kinetics. In this example, the A species is assumed to be synthesized from an effectively infinite supply of amino acids, and then gets degraded back to amino acids, producing a cycle that cannot be saturated. Inhibition of this cycle by X yields a directly proportional effect in Y. The Wnt signaling system appears to use this approach~\cite{Nunns_Goentoro_2018}. Finally, panel F shows another approach that does not use feedback, called a push-pull mechanism~\cite{Andrews_Brent_2016}. It creates linearity through symmetric signaling in which inactive forms of upstream proteins inactivate downstream proteins, and active forms activate the downstream proteins. This approach is seen in the yeast pheromone response system~\cite{Bush_Colman-Lerner_2016} and the cell-autonomous heat shock response system~\cite{Conde_Ovsenek_2009}.

\subsubsection{Hyperbolic output}

\textit{Problem.} Biochemical systems often need to be sensitive to a very wide range of input signals, while keeping the output within a finite range.

\textit{Solution.} Hyperbolic output, in which the system has linear sensitivity to low inputs, but decreasing sensitivity to higher inputs.

Hyperbolic output, or something very similar to it, is likely to be the most widely observed behavior of biochemical systems. This is because most biochemical reactions saturate with high input concentrations~\cite{Black_Leff_1983} and also, presumably, because hyperbolic outputs solve the problem of maintaining sensitivity over a wide range of inputs.

\begin{figure}[ht]
\centering
\includegraphics[height=2cm]{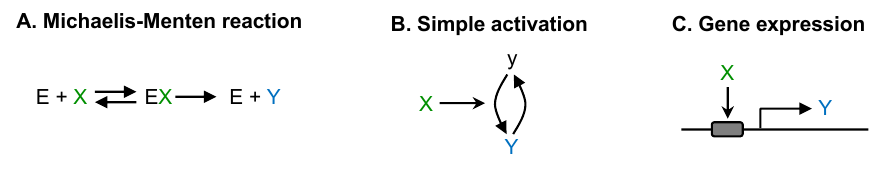}
\caption{Some mechanisms that produce the Hyperbolic output design pattern. In each panel, Y responds with hyperbolic dependence on X.}
\label{fig:HyperbolicOutput}
\end{figure}

Figure~\ref{fig:HyperbolicOutput} illustrates three mechanisms that produce hyperbolic output, again with X as the input and Y as the output. Panel A shows an enzymatically catalyzed reaction, which exhibits hyperbolic behavior due to enzyme saturation. Panel B illustrates the activation of multi-state proteins through ligand binding, phosphorylation, or some other modification; this is hyperbolic due to consumption of the inactive state. Panel C shows gene expression activation by a transcription factor, which is hyperbolic due to saturation of the transcription factor binding site. Many other simple biochemical reactions also respond linearly at low levels and saturate at high levels of input, again leading to hyperbolic or similar output. In addition, if any step within a sequence of processes has a hyperbolic output, then the entire sequence tends to exhibit hyperbolic or similar sensitivity as well~\cite{Black_Leff_1983, Andrews_Brent_2016}.

The chemotactic receptor cluster in \textit{E. coli} bacteria exhibits hyperbolic output over a remarkably wide range of input values, with both high sensitivity to low attractant concentrations and a wide dynamic range. It has been proposed that it accomplishes this through the spread of activity within the receptor cluster~\cite{Bray_Morton-Firth_1998}. Here, high activity spread between neighboring receptors causes a single ligand to activate many receptors, and hence create a large signal, but a decreased activity at high ligand concentrations still allows the cluster to respond to changes.

\subsubsection{Switching}

\textit{Problem.} Cells often need to make decisions by converting a graded input signal to a discrete output signal that is either on or off.

\textit{Solution.} Threshold detection using either ultrasensitivity or bistability.

Signaling is considered ultrasensitive if the output transitions rapidly between low and high states over a narrow range of inputs, and bistable if there are particular input values where the output can be stable in either the low or high state~\cite{Tyson_Novak_2003}. Both types of cell signaling are widespread.

Ultrasensitivity is most easily quantified through the response coefficient, which represents the relative steepness of the steady-state dose-response curve~\cite{small1990covalent,Zhang_Anderson_2013}. More precisely, most experimental dose-response curves can be described well by Hill functions~\cite{Goutelle_Maire_2008}, with the response coefficient being equal to the corresponding Hill coefficient~\cite{hofmeyr1997reversible,rohwer2010kinetic, sauro2011enzyme}, which is 1 for hyperbolic output and larger for ultrasensitive output.

\begin{figure}[ht]
\centering
\includegraphics[height=5cm]{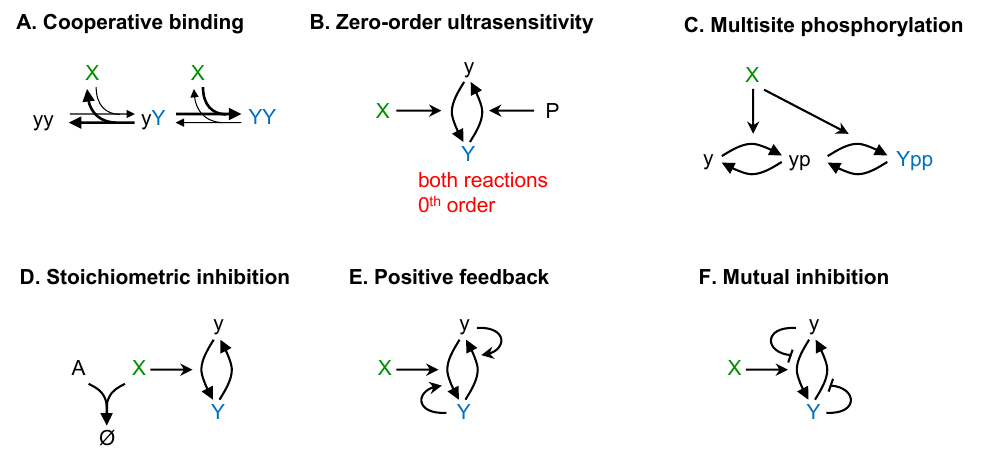}
\caption{Mechanisms for the Switching design pattern. In each panel, X is the input and Y is the output.}
\label{fig:SwitchingOutput}
\end{figure}

Figure~\ref{fig:SwitchingOutput} shows several approaches for producing ultrasensitivity~\cite{Zhang_Anderson_2013, Ferrell_Ha_2014a, Ferrell_Ha_2014b, Ferrell_Ha_2014c,straube2017analysis}. (A) Cooperative binding is a type of allostery in which the binding of one ligand increases the binding affinity for additional ligands. The classic example is oxygen binding to hemoglobin where allosteric influences between the multiple protein subunits increase the response coefficient to about 2.8. (B) In zero-order ultrasensitivity, signaling proteins that operate near saturation have rate constants that are zeroth order in their substrates. If these proteins control the activation and deactivation of two species in a covalent modification cycle, then the cycle switches abruptly between being almost entirely in one state and almost entirely in the other state~\cite{Goldbeter_Koshland_1981, Ferrell_Ha_2014a}. In practice, this effect may be limited due to enzyme sequestration~\cite{Bluthgen_Kholodenko_2006}. (C) Multi-step reactions in which the same input signal is used at multiple steps are typically ultrasensitive. For example, an input kinase is used repeatedly in multisite phosphorylation, which can cause the response coefficient to increase up to the number of phosphorylation sites \cite{Ferrell_Ha_2014b}. Linking several such events in series with kinase cascades creates even greater ultrasensitivity~\cite{Huang_Ferrell_1996}. A different example arises with dimerization and higher multimers, in which reuse of the input protein produces an ultrasensitive output signal. This could explain why many transcription factors bind DNA as dimers~\cite{Hsu_Becskei_2016}. (D) Stoichiometric inhibition uses a tight-binding inhibitor, shown as species A, that blocks the activity of an added enzyme, up until the enzyme concentration equals the inhibitor concentration. Above this threshold concentration, the enzyme starts to accumulate, which enables it to suddenly switch on downstream effects to produce an ultrasensitive response~\cite{buchler2009protein,Ferrell_Ha_2014b}. (E and F) Positive feedback or, equivalently, mutual inhibition, produces ultrasensitivity by amplifying any asymmetry in the system, driving it away from intermediate states and toward extreme states.

Bistability is an extreme version of ultrasensitivity, in which the steady-state dose-response curve goes beyond vertical to actually curve back on itself to create an S-shape. In this case, the middle inverted portion of the curve is unphysical, but the two branches that are above and below it represent the two stable states. The particular output that's observed depends on the signaling history. While all of the approaches described above are helpful for improving bistability, positive feedback is the only one that can actually create bistability. To do so, positive feedback (or mutual inhibition) increases the stability of each output state sufficiently that transitions only occur when the input is strong enough to overpower the feedback's stabilizing effect. Biological examples of bistability include the lysis and lysogeny states of phage $\lambda$ \cite{Arkin_McAdams_1998}, the clockwise and counterclockwise rotation states of \textit{E. coli} motors~\cite{Duke_Bray_2001}, a cell fate switch in \textit{Xenopus} oocytes~\cite{Ferrell_Machleder_1998}, and a synthetic transcriptional switch that was built into yeast cells~\cite{Gardner_Collins_2000}.

\subsubsection{Direction maker}

\textit{Problem.} While individual biochemical reactions are always reversible, they typically need to operate in a specific direction for cells to perform essential functions.

\textit{Solution.} Reactions can be made effectively irreversible if they either have a large free energy decrease or if they rely on a reactant that is kept at high concentration and produce a product that is kept at low concentration.

Intracellular reactions range from being sufficiently reversible that they are essentially always at equilibrium to being almost completely irreversible. This entire range is required for cells to function properly. At one extreme, reversible reactions are essential for metabolic control through product inhibition~\cite{Fell_Sauro_1985, Cornish-Bowden_Cardenas_2001}, oxygen transport via hemoglobin, and buffering mechanisms for intracellular pH control. At the other extreme, many processes need to proceed in a specific direction, such as the transcription of DNA to RNA, translation of RNA to proteins, actin polymerization at the correct filament end to create directed cell movement~\cite{Pollard_Mullins_2000}, and the metabolic biosynthesis of ATP from ADP.

From basic physical chemistry, reactions are effectively irreversible if their change in Gibbs free energy is much more negative than the available thermal energy. In practice, this occurs if reactions are very energetically favorable and/or if the reactants are much more abundant than the products. Cells use both approaches to maintain directionality. For example, directed actin dynamics occurs in several steps: ATP binds to monomeric actin proteins, they polymerize onto a filament's barbed end, their ATP hydrolyzes to ADP, and then depolymerize off the filament's pointed end. The ATP binding and hydrolysis creates irreversibility both because metabolic processes maintain several-fold greater intracellular concentrations of ATP than ADP, and because ATP hydrolysis is energetically favorable. Together, these cause actin filaments to ``treadmill'' in a specific direction, with constant growth at one end and disassembly at the other. All other cellular processes that include ATP hydrolysis are essentially irreversible for the same reasons.

Unidirectional information transfer in kinase cascades appears different at first glance, but is also maintained through ATP binding and hydrolysis. Here, each kinase binds ATP and then hydrolyzes it to ADP in order to phosphorylate its target. Nevertheless, there can be some information transfer to upstream elements even in kinase cascades. This retroactivity occurs due to kinase sequestration by downstream elements, which then feeds back upstream through the effect on free kinase availability~\cite{Ventura_Merajver_2008, Ventura_Ninfa_2010}.

Proteolysis represents a different type of irreversible reaction. Its large negative free energy arises from the large entropic increase that results from disassembling an ordered protein. Regulated proteolysis has been shown to be an important step in cell development, enabling irreversible commitment to cell fate decisions~\cite{Gottesman_1999, Jenal_Hengge-Aronis_2003, Lai_2004}.

\subsubsection{Insulator}

\textit{Problem.} Cellular reaction networks are highly interconnected but require modularization to reduce undesirable crosstalk and improve evolvability.

\textit{Solution.} Create module boundaries that insulate subnetworks from each other.

Intracellular biochemical reaction networks have been shown to exhibit substantial modularity~\cite{Hartwell_Murray_1999, Guimera_Amaral_2005}. This appears to arise from evolutionary selection pressure that favors lower connection costs within cells \cite{Clune_Lipson_2013} and also the benefits that modularity provides for improving the ability of an organism to adapt to a changing environment through greater network evolvability. As in human-engineered systems, modular biochemical systems are more adaptable because mutations to module components only affect behavior within the module, while leaving the rest of the system largely unchanged.

For a module to act independently of the rest of the system, it needs to accept information or material through its inputs without substantially perturbing upstream processes. Conversely, its outputs need to be sufficiently robust that downstream processes don't affect them, and hence don't affect other dynamics within the module. In electronics terminology, a module needs a high input impedance, meaning that it draws little current from upstream processes, and a low output impedance, meaning that it can supply large amounts of current to downstream processes.

Modules can be insulated from their surrounding systems through several mechanisms. (1) Spatial localization maintains strong internal connections while reducing external ones. For example, the many protein translation steps all occur within ribosomes and are often followed by direct transfer to chaperone proteins for folding~\cite{Preissler_Deuerling_2012}. Metabolism is likely to occur largely through metabolic channeling in which substrates are passed directly from one enzyme to another~\cite{Pareek_Benkovic_2021}, and much signal processing occurs in large multi-protein complexes~\cite{Langeberg_Scott_2015}. (2) Standardized connections enable multiple modules to interact with the external system without perturbing each other~\cite{Csete_Doyle_2002}. For example, eukaryotic cells have many signaling systems, each of which represents a separate module, and each of which terminates at the activation of a transcription factor, which is a standardized connection. Likewise, catabolic metabolism creates ATP as a standard connection, which insulates these metabolic processes from other cellular processes. (3) For signaling pathways, the input end of the module can be insulated from upstream processes using a large amplification~\cite{DelVecchio_Sontag_2008}. For example, both transcription from a strong promoter and phosphorylation by a kinase protein can create large signal amplifications with minimal retroactivity. (4) At the output end of modules, negative feedback is widely used to keep the output constant, independent of the impact of downstream processes~\cite{Hofmeyr_Cornish-Bowden_2000, DelVecchio_Sontag_2008}. Examples can be found in amino acid biosynthesis, where negative feedback is used to isolate biosynthesis of amino acids from protein synthesis~\cite{curien2009understanding}. Other instances of homeostasis also promote modularity in the same manner. (5) In signaling systems, covalent cycles that have fast turnover have been shown to help insulate modules because they are able to supply adequate signaling protein for downstream processes while being minimally susceptible to retroactivity due to the separation of timescales~\cite{DelVecchio_Sontag_2008, Franco_Murray_2009, Shah_DelVecchio_2016}.


\subsubsection{Fold-change and ratiometric detection}

\textit{Problem.} Cells often need to act on chemical concentrations with quantitative accuracy, despite having variable numbers of proteins that sense that chemical.

\textit{Solution.} Use reaction networks that compare the number of ligand-bound proteins either over time or against the number of unbound proteins.

Intracellular protein concentrations vary substantially between different individual cells, even if they are isogenic~\cite{Elowitz_Swain_2002}, and also vary over time for a single cell. Despite this variation, cells need to measure ligand concentrations with good quantitative accuracy so that they can make well-informed decisions, whether for chemotaxis, mating, apoptosis, or other outcomes. Two approaches for addressing this problem are fold-change detection, in which a cell determines the percent change in protein binding over time, and ratiometric detection, in which a cell determines the fraction of those proteins that are bound to ligands.

Fold-change detection derives from Weber's law, which expresses the observation that the minimum detectable difference in a stimulus tends to be directly proportional to the total value of the stimulus~\cite{Ferrell_2009}. It also relates to the observation that the stochastic noise of protein expression scales with protein abundance~\cite{Bar-Even_Barkai_2006}. Fold-change detection extends these ideas to the concept that all aspects of a cell's response, including both amplitude and temporal effects, depend only on fold changes of the input and not on its absolute levels~\cite{Shoval_Alon_2010}. Fold-change detection is widely observed, including in \textit{E. coli} chemotaxis~\cite{Webre_Stock_2003, Tu_Berg_2008, Shoval_Alon_2010, Clausznitzer_Endres_2014} and the NF-$\kappa$B~\cite{Lee_Gaudet_2014}, EGF~\cite{Cohen-Saidon_Alon_2009}, and Wnt signaling \cite{Goentoro_Kirschner_2009} systems. An intriguing benefit of fold-change detection is that it enables cells to compare signals from multiple types of inputs because each gets quantified relative to its own prior level~\cite{Hart_Alon_2013b}. Fold-change detection requires a memory of the signal over time so that the cell can compare the new signal to its prior level. This can be achieved with an incoherent feedforward motif~\cite{Goentoro_Alon_2009}, in which an input signal activates a downstream element and then, after a short delay, represses the same downstream element. The activation and repression are directly proportional to each other, causing them to cancel out at steady-state and also, if parameters are in the appropriate ranges, making the output respond to the fold change in the input. Other mechanisms include nonlinear integral feedback and logarithmic input with linear feedback \cite{Shoval_Alon_2010}.

In ratiometric sensing, a cell determines the fraction of proteins that are activated at a specific time by comparing signals from both the active and inactive forms. This can be accomplished through the push-pull mechanism described above and shown in Figure~\ref{fig:ProportionalOutput}F~\cite{Andrews_Brent_2016}. It is a form of paradoxical signaling, defined as a component simultaneously acting in two opposing ways on its target~\cite{Shinar_Alon_2007, Hart_Alon_2013a}. The symmetric behavior of the push-pull mechanism makes it sensitive to the fraction of activated protein rather than the absolute amount. It appears to be widely used, including in yeast pheromone response signaling \cite{Bush_Colman-Lerner_2016, Henderson_Lew_2019}, heat shock response~\cite{Conde_Ovsenek_2009}, and the \textit{E. coli} EnvZ-OmpR system~\cite{Russo_Silhavy_1993, Batchelor_Goulian_2003}.








\subsection{Behavioral patterns in the yeast pheromone response system}

\begin{figure}[ht]
\centering
\includegraphics[height=6cm]{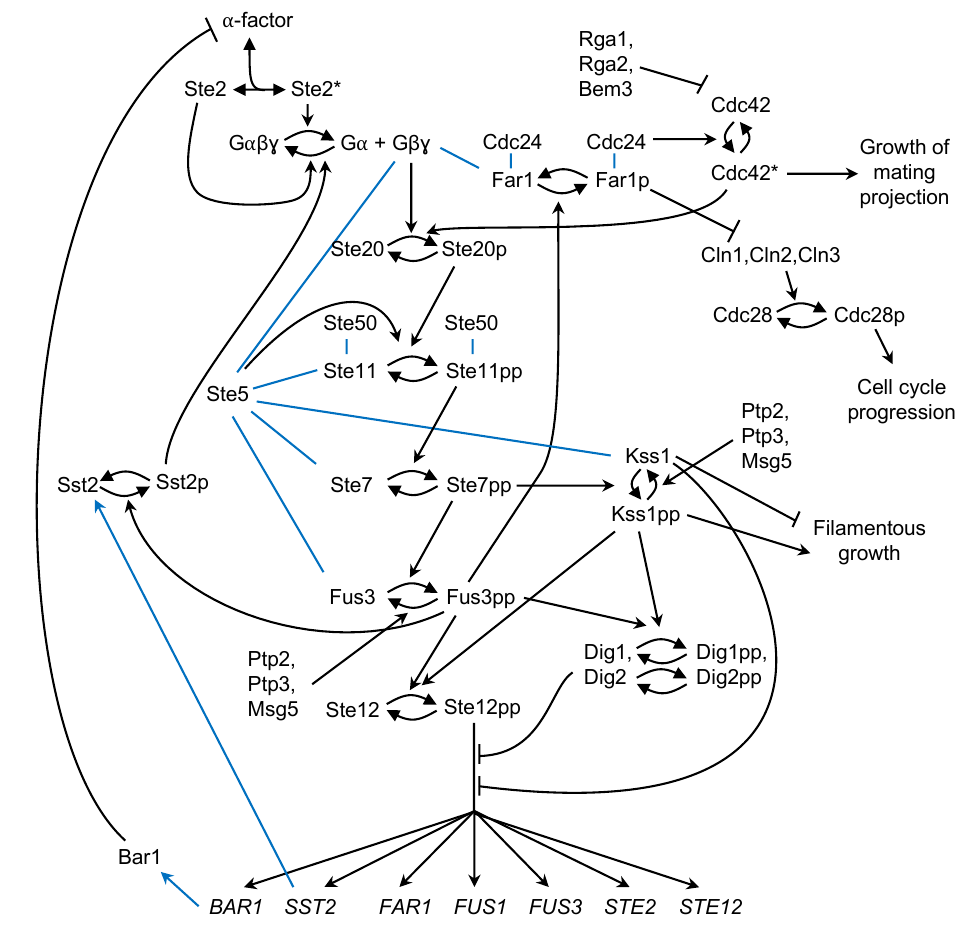}
\caption{Diagram of the yeast signaling reaction network. Black arrows with solid head represent chemical reactions, black arrows with barbed and T-bar points represent activations and inhibitions, blue arrows represent gene expression, and blue lines represent binding.}
\label{fig:YeastSignaling}
\end{figure}

A benefit of viewing a chemical reaction network through the lens of design patterns is that it encourages abstraction of the network details into generalized behaviors. We illustrate this abstraction with the yeast pheromone response system (Figure~\ref{fig:YeastSignaling}), which is exceptionally well understood~\cite{Kurjan_1992, Wang_Dohlman_2004, Bardwell_2005, Bush_Colman-Lerner_2016}.

In this system, signaling starts when $\alpha$-factor (pheromone) binds to the Ste2 G-protein coupled receptor (GPCR). This promotes dissociation of the heterotrimeric G-protein, G$\alpha \beta \gamma$ to separate G$\alpha$ and G$\beta \gamma$. The later portion helps activate Ste20 through phosphorylation. In conjunction with the Ste5 scaffold protein, Ste20 then sends the signal through a MAP kinase cascade that starts with double phosphorylation of Ste11, then double phosphorylation of Ste7, and then double phosphorylation of the MAP kinases Fus3 and Kss1. Both of these MAP kinases promote activation of the Ste12 transcription factor, of which Fus3 acts primarily by activating Ste12 directly and Kss1 acts primarily by repressing Dig1 and Dig2, which then relieves their repression of Ste12. Ste12 then activates transcription of the pheromone responsive genes. Several other pathways branch off this one. The first branch drives growth of the yeast cell mating projection, a second arrests progression through the cell cycle in preparation for mating, and a third promotes filamentous growth \cite{Song_Kumar_2012}.

\begin{figure}[ht]
\centering
\includegraphics[height=5cm]{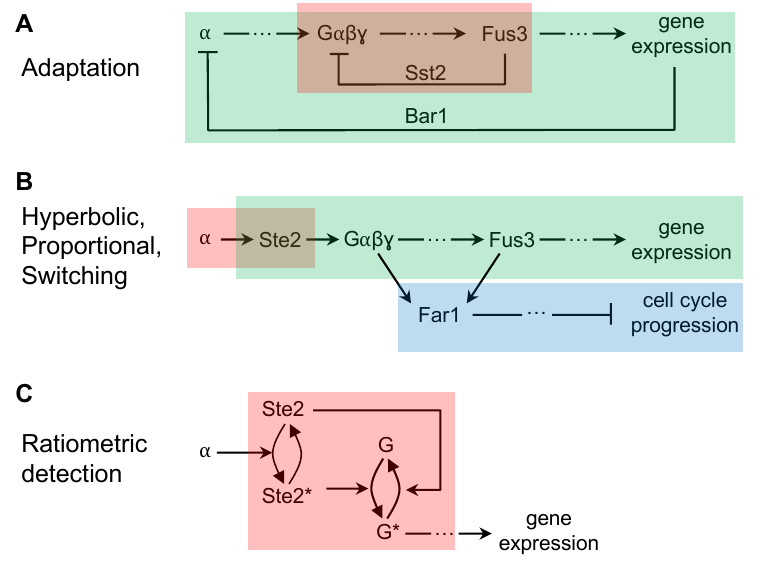}
\caption{Several behavioral patterns in the yeast pheromone response signaling reaction network. (A) Both boxes represent the Adaptation pattern. (B) The red box represents Hyperbolic output, green represents Proportional output, and blue represents Switching output. (C) The box represents Ratiometric detection.}
\label{fig:BehavioralPatternsYeast}
\end{figure}

This system adapts to pheromone over time, which represents the Adaptation design pattern. It works through multiple negative feedback loops (Figure~\ref{fig:BehavioralPatternsYeast}A). In one, the Sst2 protein forms a negative feedback by promoting GTP hydrolysis in the G$\alpha$ subunit of the G-protein, leading to G-protein association and inactivation. This helps reset the system after pheromone stimulation, decreasing pheromone-induced expression and returning the cell to normal growth~\cite{Dohlman_Thorner_1996}. Additionally, not shown in the figure, Sst2 feedback helps reset the system by promoting receptor (Ste2) recovery~\cite{Venkatapurapu_Elston_2015} and increasing cyclin activity \cite{Chen_Kurjan_1997}. On longer timescales, Bar1 expression and secretion degrades extracellular pheromone, also causing system adaptation~\cite{Jin_Elston_2011}. The \textit{BAR1} gene is often deleted in signaling research for this reason, enabling temporally stable signaling~\cite{Yu_Brent_2008}. These mechanisms do not use integral feedback control, so they create partial but not perfect adaptation.

Focusing on the steady-state dose-response relationships in the system, some are proportional, others are hyperbolic, and yet others are ultrasensitive (Figure~\ref{fig:BehavioralPatternsYeast}B). The Proportional output pattern is demonstrated by the levels of receptor binding, G-protein dissociation, Fus3 phosphorylation, and gene expression being strongly correlated with each other~\cite{Yi_Simon_2003, Yu_Brent_2008}. This has been termed DoRA, for dose-response alignment, and indicates proportional signaling between each of these steps, and also proportional signaling over the entire pathway \cite{Poritz_Kamb_2001}. The system presumably exhibits this pattern to improve information transmission through the system~\cite{Yu_Brent_2008}. The Hyperbolic output pattern appears in pheromone binding to the Ste2 cell-surface receptors \cite{Jenness_Hartwell_1986, Bajaj_Dumont_2004} and is undoubtedly present for many of the enzymatic reactions in the system~\cite{Kofahl_Klipp_2004, Suderman_Deeds_2013}. Gene expression may be hyperbolic in this system as well, although it's noteworthy that the Ste12 transcription factor likely binds DNA in dimeric form~\cite{Olson_Sadowski_2000}, which may create weak ultrasensitivity. The Switching output pattern is observed in the two pathway branches that do not lead to transcription, which are the filamentous growth pathway~\cite{Cook_Thorner_1997} and a pathway that leads to cell cycle arrest~\cite{Hao_Dohlman_2008}.

Finally, this system's sensitivity to pheromone has been shown to be independent of its number of Ste2 receptors~\cite{Bush_Colman-Lerner_2016}. This is an example of the Ratiometric detection design pattern, which is accomplished with a push-pull mechanism at the start of the signaling pathway (Figure~\ref{fig:BehavioralPatternsYeast}C). Here, receptor-bound Ste2 promotes the dissociation of G-protein, and receptor-unbound Ste2 actively promotes the association of G-protein \cite{Bush_Colman-Lerner_2016}.

\section{Discussion}

To repeat the definition given above, design patterns are recurrent solutions to commonly encountered problems. All biological cells encounter the same problems of how to construct the biochemical components that they are built from, how to connect those components together into useful reaction networks, and how to use those reaction networks to animate life. The generalized solutions to these problems are the constructive, structural, and behavioral design patterns listed here.

The idea of understanding cellular systems in terms of functional parts is of course not new~\cite{Bhalla_2003,sauro2004quantitative,lim2013design}. However, we present a larger scope and consider a different perspective. This design pattern view is useful for abstracting a broad range of cell functions into a manageable set of distinct patterns, enabling one to better see parallels and differences between different cell systems. It also helps build an understanding of what tools cells have to work with, and why different cellular mechanisms operate as they do.

The design pattern concept can also be usefully applied using a narrower scope. For example, Bhalla described the proteins in a signaling network through their functions as multipliers, adders, integrators, and delayers~\cite{Bhalla_2003}, all of which could be seen as separate types of design patterns. Also, we presented many types of mechanisms for several of the behavioral patterns listed above, each of which could be interpreted as its own design pattern. We plan to apply this narrower view to future work in which we will consider the deconstruction of the EGFR/ERK signaling pathway into a series of interlinked and hierarchical design patterns, allowing us to better understand this important signaling network that impacts many disease states. In turn, a better understanding is likely to lead to new ways to think about therapeutic interventions.

A popular thought experiment is to consider what would happen if the ``tape of life'' could be rewound and started again, perhaps 2 billion years ago, and to see how life would evolve in that alternate universe~\cite{Gould_1989}. In many ways, the outcome would presumably be very different from the biology that we see around us today due to cumulative impacts of stochastic events. On the other hand, those alternate biological cells would almost certainly face many of the same problems that today's cells faced in their actual evolution, and would likely evolve many of the same solutions. Going even farther afield, one can speculate about life on other planets, where again the same problems would likely arise, and would probably be addressed with the same solutions. If so, then the design patterns listed here, along with others not addressed, could be reasonably considered universal principles of life.

This work connects back to the computer science origins of design patterns in several interesting ways. First, there is substantial current interest in building whole cell computer simulations~\cite{Tomita_2001, Karr_Covert_2012}, leading to the question of what capabilities those simulators would need to have. One answer is that those simulators would need to address the same problems as actual biological cells, and thus would require the same solutions. As a result, this catalog of biological design patterns can also be considered a list of the computational design patterns that need to be developed to enable whole-cell simulations. Second, computer science methods could be harnessed to better identify biological design patterns. In particular, it may be possible to train modern AI/ML methods to identify design patterns automatically in a given network, from which they could then posit specific interventions to change network behavior.

\backmatter

\bmhead{Acknowledgments}

We thank Cong Trinh for permission to republish their figure on \textit{E. coli} metabolism, Song Feng and Michael Kochen for useful discussions. The research reported in this article was supported by the National Cancer Institute of the National Institutes of Health under award number U01CA227544. The authors declare that they have no conflicts of interest.

\bibliography{DesignPatternsRefs}


\begin{thebibliography}{174}
\ifx \bisbn   \undefined \def \bisbn  #1{ISBN #1}\fi
\ifx \binits  \undefined \def \binits#1{#1}\fi
\ifx \bauthor  \undefined \def \bauthor#1{#1}\fi
\ifx \batitle  \undefined \def \batitle#1{#1}\fi
\ifx \bjtitle  \undefined \def \bjtitle#1{#1}\fi
\ifx \bvolume  \undefined \def \bvolume#1{\textbf{#1}}\fi
\ifx \byear  \undefined \def \byear#1{#1}\fi
\ifx \bissue  \undefined \def \bissue#1{#1}\fi
\ifx \bfpage  \undefined \def \bfpage#1{#1}\fi
\ifx \blpage  \undefined \def \blpage #1{#1}\fi
\ifx \burl  \undefined \def \burl#1{\textsf{#1}}\fi
\ifx \doiurl  \undefined \def \doiurl#1{\url{https://doi.org/#1}}\fi
\ifx \betal  \undefined \def \betal{\textit{et al.}}\fi
\ifx \binstitute  \undefined \def \binstitute#1{#1}\fi
\ifx \binstitutionaled  \undefined \def \binstitutionaled#1{#1}\fi
\ifx \bctitle  \undefined \def \bctitle#1{#1}\fi
\ifx \beditor  \undefined \def \beditor#1{#1}\fi
\ifx \bpublisher  \undefined \def \bpublisher#1{#1}\fi
\ifx \bbtitle  \undefined \def \bbtitle#1{#1}\fi
\ifx \bedition  \undefined \def \bedition#1{#1}\fi
\ifx \bseriesno  \undefined \def \bseriesno#1{#1}\fi
\ifx \blocation  \undefined \def \blocation#1{#1}\fi
\ifx \bsertitle  \undefined \def \bsertitle#1{#1}\fi
\ifx \bsnm \undefined \def \bsnm#1{#1}\fi
\ifx \bsuffix \undefined \def \bsuffix#1{#1}\fi
\ifx \bparticle \undefined \def \bparticle#1{#1}\fi
\ifx \barticle \undefined \def \barticle#1{#1}\fi
\bibcommenthead
\ifx \bconfdate \undefined \def \bconfdate #1{#1}\fi
\ifx \botherref \undefined \def \botherref #1{#1}\fi
\ifx \url \undefined \def \url#1{\textsf{#1}}\fi
\ifx \bchapter \undefined \def \bchapter#1{#1}\fi
\ifx \bbook \undefined \def \bbook#1{#1}\fi
\ifx \bcomment \undefined \def \bcomment#1{#1}\fi
\ifx \oauthor \undefined \def \oauthor#1{#1}\fi
\ifx \citeauthoryear \undefined \def \citeauthoryear#1{#1}\fi
\ifx \endbibitem  \undefined \def \endbibitem {}\fi
\ifx \bconflocation  \undefined \def \bconflocation#1{#1}\fi
\ifx \arxivurl  \undefined \def \arxivurl#1{\textsf{#1}}\fi
\csname PreBibitemsHook\endcsname

\bibitem{Alexander_1966}
\begin{barticle}
\bauthor{\bsnm{Alexander}, \binits{C.}}:
\batitle{The pattern of streets}.
\bjtitle{Journal of the American Institute of Planners}
\bvolume{32}(\bissue{5}),
\bfpage{273}--\blpage{278}
(\byear{1966})
\end{barticle}
\endbibitem

\bibitem{Alexander_1977}
\begin{bbook}
\bauthor{\bsnm{Alexander}, \binits{C.}},
\bauthor{\bsnm{Ishikawa}, \binits{S.}},
\bauthor{\bsnm{Silverstein}, \binits{M.}},
\bauthor{\bsnm{Jacobson}, \binits{M.}},
\bauthor{\bsnm{Fiksdahl-King}, \binits{I.}},
\bauthor{\bsnm{Angel}, \binits{S.}}:
\bbtitle{A Pattern Language: Towns, Buildings, Construction}.
\bpublisher{Oxford university press},
\blocation{New York}
(\byear{1977})
\end{bbook}
\endbibitem

\bibitem{Alexander_1979}
\begin{bbook}
\bauthor{\bsnm{Alexander}, \binits{C.}}:
\bbtitle{The Timeless Way of Building}.
\bpublisher{Oxford university press},
\blocation{New York}
(\byear{1979})
\end{bbook}
\endbibitem

\bibitem{Gamma_Vlissides_1995}
\begin{bbook}
\bauthor{\bsnm{Gamma}, \binits{E.}},
\bauthor{\bsnm{Johnson}, \binits{R.}},
\bauthor{\bsnm{Helm}, \binits{R.}},
\bauthor{\bsnm{Johnson}, \binits{R.E.}},
\bauthor{\bsnm{Vlissides}, \binits{J.}}:
\bbtitle{Design Patterns: Elements of Reusable Object-oriented Software}.
\bpublisher{Addison-Wesley},
\blocation{New York}
(\byear{1995})
\end{bbook}
\endbibitem

\bibitem{McAdams_Shapiro_1995}
\begin{barticle}
\bauthor{\bsnm{McAdams}, \binits{H.H.}},
\bauthor{\bsnm{Shapiro}, \binits{L.}}:
\batitle{Circuit simulation of genetic networks}.
\bjtitle{Science}
\bvolume{269}(\bissue{5224}),
\bfpage{650}--\blpage{656}
(\byear{1995})
\end{barticle}
\endbibitem

\bibitem{Hartwell_Murray_1999}
\begin{barticle}
\bauthor{\bsnm{Hartwell}, \binits{L.H.}},
\bauthor{\bsnm{Hopfield}, \binits{J.J.}},
\bauthor{\bsnm{Leibler}, \binits{S.}},
\bauthor{\bsnm{Murray}, \binits{A.W.}}:
\batitle{From molecular to modular cell biology}.
\bjtitle{Nature}
\bvolume{402}(\bissue{Suppl 6761}),
\bfpage{47}--\blpage{52}
(\byear{1999})
\end{barticle}
\endbibitem

\bibitem{Lazebnik_2002}
\begin{barticle}
\bauthor{\bsnm{Lazebnik}, \binits{Y.}}:
\batitle{Can a biologist fix a radio?---or, what {I} learned while studying
  apoptosis}.
\bjtitle{Cancer cell}
\bvolume{2}(\bissue{3}),
\bfpage{179}--\blpage{182}
(\byear{2002})
\end{barticle}
\endbibitem

\bibitem{Bhalla_2003}
\begin{barticle}
\bauthor{\bsnm{Bhalla}, \binits{U.S.}}:
\batitle{Understanding complex signaling networks through models and
  metaphors}.
\bjtitle{Progress in biophysics and molecular biology}
\bvolume{81}(\bissue{1}),
\bfpage{45}--\blpage{65}
(\byear{2003})
\end{barticle}
\endbibitem

\bibitem{Sauro_Kholodenko_2004}
\begin{barticle}
\bauthor{\bsnm{Sauro}, \binits{H.M.}},
\bauthor{\bsnm{Kholodenko}, \binits{B.N.}}:
\batitle{Quantitative analysis of signaling networks}.
\bjtitle{Progress in biophysics and molecular biology}
\bvolume{86}(\bissue{1}),
\bfpage{5}--\blpage{43}
(\byear{2004})
\end{barticle}
\endbibitem

\bibitem{Bray_2009}
\begin{bbook}
\bauthor{\bsnm{Bray}, \binits{D.}}:
\bbtitle{Wetware: {A} Computer in Every Living Cell}.
\bpublisher{Yale University Press},
\blocation{Yale}
(\byear{2009})
\end{bbook}
\endbibitem

\bibitem{Milo_Alon_2002}
\begin{barticle}
\bauthor{\bsnm{Milo}, \binits{R.}},
\bauthor{\bsnm{Shen-Orr}, \binits{S.}},
\bauthor{\bsnm{Itzkovitz}, \binits{S.}},
\bauthor{\bsnm{Kashtan}, \binits{N.}},
\bauthor{\bsnm{Chklovskii}, \binits{D.}},
\bauthor{\bsnm{Alon}, \binits{U.}}:
\batitle{Network motifs: simple building blocks of complex networks}.
\bjtitle{Science}
\bvolume{298}(\bissue{5594}),
\bfpage{824}--\blpage{827}
(\byear{2002})
\end{barticle}
\endbibitem

\bibitem{Milo_Alon_2004}
\begin{barticle}
\bauthor{\bsnm{Milo}, \binits{R.}},
\bauthor{\bsnm{Itzkovitz}, \binits{S.}},
\bauthor{\bsnm{Kashtan}, \binits{N.}},
\bauthor{\bsnm{Levitt}, \binits{R.}},
\bauthor{\bsnm{Shen-Orr}, \binits{S.}},
\bauthor{\bsnm{Ayzenshtat}, \binits{I.}},
\bauthor{\bsnm{Sheffer}, \binits{M.}},
\bauthor{\bsnm{Alon}, \binits{U.}}:
\batitle{Superfamilies of evolved and designed networks}.
\bjtitle{Science}
\bvolume{303}(\bissue{5663}),
\bfpage{1538}--\blpage{1542}
(\byear{2004})
\end{barticle}
\endbibitem

\bibitem{Tyson_Novak_2003}
\begin{barticle}
\bauthor{\bsnm{Tyson}, \binits{J.J.}},
\bauthor{\bsnm{Chen}, \binits{K.C.}},
\bauthor{\bsnm{Novak}, \binits{B.}}:
\batitle{Sniffers, buzzers, toggles and blinkers: dynamics of regulatory and
  signaling pathways in the cell}.
\bjtitle{Current opinion in cell biology}
\bvolume{15}(\bissue{2}),
\bfpage{221}--\blpage{231}
(\byear{2003})
\end{barticle}
\endbibitem

\bibitem{DelVecchio_Sontag_2008}
\begin{barticle}
\bauthor{\bsnm{Del~Vecchio}, \binits{D.}},
\bauthor{\bsnm{Ninfa}, \binits{A.J.}},
\bauthor{\bsnm{Sontag}, \binits{E.D.}}:
\batitle{Modular cell biology: retroactivity and insulation}.
\bjtitle{Molecular systems biology}
\bvolume{4}(\bissue{1}),
\bfpage{161}
(\byear{2008})
\end{barticle}
\endbibitem

\bibitem{Sauro_2008}
\begin{barticle}
\bauthor{\bsnm{Sauro}, \binits{H.M.}}:
\batitle{Modularity defined}.
\bjtitle{Molecular systems biology}
\bvolume{4}(\bissue{1}),
\bfpage{166}
(\byear{2008})
\end{barticle}
\endbibitem

\bibitem{hopfield1974kinetic}
\begin{barticle}
\bauthor{\bsnm{Hopfield}, \binits{J.J.}}:
\batitle{Kinetic proofreading: a new mechanism for reducing errors in
  biosynthetic processes requiring high specificity}.
\bjtitle{Proceedings of the National Academy of Sciences}
\bvolume{71}(\bissue{10}),
\bfpage{4135}--\blpage{4139}
(\byear{1974})
\end{barticle}
\endbibitem

\bibitem{Boeger_2022}
\begin{barticle}
\bauthor{\bsnm{Boeger}, \binits{H.}}:
\batitle{Kinetic proofreading}.
\bjtitle{Annual Review of Biochemistry}
\bvolume{91},
\bfpage{423}--\blpage{447}
(\byear{2022})
\end{barticle}
\endbibitem

\bibitem{Hofmeyr_Cornish-Bowden_2000}
\begin{barticle}
\bauthor{\bsnm{Hofmeyr}, \binits{J.-H.S.}},
\bauthor{\bsnm{Cornish-Bowden}, \binits{A.}}:
\batitle{Regulating the cellular economy of supply and demand}.
\bjtitle{FEBS letters}
\bvolume{476}(\bissue{1-2}),
\bfpage{47}--\blpage{51}
(\byear{2000})
\end{barticle}
\endbibitem

\bibitem{Emiola_Andrews_2015}
\begin{barticle}
\bauthor{\bsnm{Emiola}, \binits{A.}},
\bauthor{\bsnm{George}, \binits{J.}},
\bauthor{\bsnm{Andrews}, \binits{S.S.}}:
\batitle{A complete pathway model for lipid a biosynthesis in escherichia
  coli}.
\bjtitle{PloS one}
\bvolume{10}(\bissue{4}),
\bfpage{0121216}
(\byear{2015})
\end{barticle}
\endbibitem

\bibitem{curien2009understanding}
\begin{barticle}
\bauthor{\bsnm{Curien}, \binits{G.}},
\bauthor{\bsnm{Bastien}, \binits{O.}},
\bauthor{\bsnm{Robert-Genthon}, \binits{M.}},
\bauthor{\bsnm{Cornish-Bowden}, \binits{A.}},
\bauthor{\bsnm{C{\'a}rdenas}, \binits{M.L.}},
\bauthor{\bsnm{Dumas}, \binits{R.}}:
\batitle{Understanding the regulation of aspartate metabolism using a model
  based on measured kinetic parameters}.
\bjtitle{Molecular systems biology}
\bvolume{5}(\bissue{1}),
\bfpage{271}
(\byear{2009})
\end{barticle}
\endbibitem

\bibitem{Anfinsen_1973}
\begin{barticle}
\bauthor{\bsnm{Anfinsen}, \binits{C.B.}}:
\batitle{Principles that govern the folding of protein chains}.
\bjtitle{Science}
\bvolume{181}(\bissue{4096}),
\bfpage{223}--\blpage{230}
(\byear{1973})
\end{barticle}
\endbibitem

\bibitem{Marianayagam_Matthews_2004}
\begin{barticle}
\bauthor{\bsnm{Marianayagam}, \binits{N.J.}},
\bauthor{\bsnm{Sunde}, \binits{M.}},
\bauthor{\bsnm{Matthews}, \binits{J.M.}}:
\batitle{The power of two: protein dimerization in biology}.
\bjtitle{Trends in biochemical sciences}
\bvolume{29}(\bissue{11}),
\bfpage{618}--\blpage{625}
(\byear{2004})
\end{barticle}
\endbibitem

\bibitem{Rivas_Minton_2001}
\begin{barticle}
\bauthor{\bsnm{Rivas}, \binits{G.}},
\bauthor{\bsnm{Fern{\'a}ndez}, \binits{J.A.}},
\bauthor{\bsnm{Minton}, \binits{A.P.}}:
\batitle{Direct observation of the enhancement of noncooperative protein
  self-assembly by macromolecular crowding: indefinite linear self-association
  of bacterial cell division protein ftsz}.
\bjtitle{Proceedings of the National Academy of Sciences}
\bvolume{98}(\bissue{6}),
\bfpage{3150}--\blpage{3155}
(\byear{2001})
\end{barticle}
\endbibitem

\bibitem{Gardner_Odde_2011}
\begin{barticle}
\bauthor{\bsnm{Gardner}, \binits{M.K.}},
\bauthor{\bsnm{Charlebois}, \binits{B.D.}},
\bauthor{\bsnm{J{\'a}nosi}, \binits{I.M.}},
\bauthor{\bsnm{Howard}, \binits{J.}},
\bauthor{\bsnm{Hunt}, \binits{A.J.}},
\bauthor{\bsnm{Odde}, \binits{D.J.}}:
\batitle{Rapid microtubule self-assembly kinetics}.
\bjtitle{Cell}
\bvolume{146}(\bissue{4}),
\bfpage{582}--\blpage{592}
(\byear{2011})
\end{barticle}
\endbibitem

\bibitem{Goldman_Bray_2004}
\begin{barticle}
\bauthor{\bsnm{Goldman}, \binits{J.}},
\bauthor{\bsnm{Andrews}, \binits{S.}},
\bauthor{\bsnm{Bray}, \binits{D.}}:
\batitle{Size and composition of membrane protein clusters predicted by monte
  carlo analysis}.
\bjtitle{European Biophysics Journal}
\bvolume{33},
\bfpage{506}--\blpage{512}
(\byear{2004})
\end{barticle}
\endbibitem

\bibitem{Mayer_Yu_2018}
\begin{barticle}
\bauthor{\bsnm{Mayer}, \binits{B.J.}},
\bauthor{\bsnm{Yu}, \binits{J.}}:
\batitle{Protein clusters in phosphotyrosine signal transduction}.
\bjtitle{Journal of molecular biology}
\bvolume{430}(\bissue{22}),
\bfpage{4547}--\blpage{4556}
(\byear{2018})
\end{barticle}
\endbibitem

\bibitem{Brangwynne_Hyman_2009}
\begin{barticle}
\bauthor{\bsnm{Brangwynne}, \binits{C.P.}},
\bauthor{\bsnm{Eckmann}, \binits{C.R.}},
\bauthor{\bsnm{Courson}, \binits{D.S.}},
\bauthor{\bsnm{Rybarska}, \binits{A.}},
\bauthor{\bsnm{Hoege}, \binits{C.}},
\bauthor{\bsnm{Gharakhani}, \binits{J.}},
\bauthor{\bsnm{J{\"u}licher}, \binits{F.}},
\bauthor{\bsnm{Hyman}, \binits{A.A.}}:
\batitle{Germline p granules are liquid droplets that localize by controlled
  dissolution/condensation}.
\bjtitle{Science}
\bvolume{324}(\bissue{5935}),
\bfpage{1729}--\blpage{1732}
(\byear{2009})
\end{barticle}
\endbibitem

\bibitem{Simons_Sampaio_2011}
\begin{barticle}
\bauthor{\bsnm{Simons}, \binits{K.}},
\bauthor{\bsnm{Sampaio}, \binits{J.L.}}:
\batitle{Membrane organization and lipid rafts}.
\bjtitle{Cold Spring Harbor perspectives in biology}
\bvolume{3}(\bissue{10}),
\bfpage{004697}
(\byear{2011})
\end{barticle}
\endbibitem

\bibitem{Schmidt_Lipkow_2014}
\begin{barticle}
\bauthor{\bsnm{Schmidt}, \binits{H.G.}},
\bauthor{\bsnm{Sewitz}, \binits{S.}},
\bauthor{\bsnm{Andrews}, \binits{S.S.}},
\bauthor{\bsnm{Lipkow}, \binits{K.}}:
\batitle{An integrated model of transcription factor diffusion shows the
  importance of intersegmental transfer and quaternary protein structure for
  target site finding}.
\bjtitle{PLOS one}
\bvolume{9}(\bissue{10}),
\bfpage{108575}
(\byear{2014})
\end{barticle}
\endbibitem

\bibitem{Sorokina_Koonin_2022}
\begin{barticle}
\bauthor{\bsnm{Sorokina}, \binits{I.}},
\bauthor{\bsnm{Mushegian}, \binits{A.R.}},
\bauthor{\bsnm{Koonin}, \binits{E.V.}}:
\batitle{Is protein folding a thermodynamically unfavorable, active,
  energy-dependent process?}
\bjtitle{International Journal of Molecular Sciences}
\bvolume{23}(\bissue{1}),
\bfpage{521}
(\byear{2022})
\end{barticle}
\endbibitem

\bibitem{Ghosh_Stuehr_2011}
\begin{barticle}
\bauthor{\bsnm{Ghosh}, \binits{A.}},
\bauthor{\bsnm{Chawla-Sarkar}, \binits{M.}},
\bauthor{\bsnm{Stuehr}, \binits{D.J.}}:
\batitle{Hsp90 interacts with inducible no synthase client protein in its
  heme-free state and then drives heme insertion by an atp-dependent process}.
\bjtitle{The FASEB Journal}
\bvolume{25}(\bissue{6}),
\bfpage{2049}
(\byear{2011})
\end{barticle}
\endbibitem

\bibitem{Tupta_Ghosh_2022}
\begin{barticle}
\bauthor{\bsnm{Tupta}, \binits{B.}},
\bauthor{\bsnm{Stuehr}, \binits{E.}},
\bauthor{\bsnm{Sumi}, \binits{M.P.}},
\bauthor{\bsnm{Sweeny}, \binits{E.A.}},
\bauthor{\bsnm{Smith}, \binits{B.}},
\bauthor{\bsnm{Stuehr}, \binits{D.J.}},
\bauthor{\bsnm{Ghosh}, \binits{A.}}:
\batitle{Gapdh is involved in the heme-maturation of myoglobin and hemoglobin}.
\bjtitle{FASEB journal: official publication of the Federation of American
  Societies for Experimental Biology}
\bvolume{36}(\bissue{2}),
\bfpage{22099}
(\byear{2022})
\end{barticle}
\endbibitem

\bibitem{Howard_2001}
\begin{bbook}
\bauthor{\bsnm{Howard}, \binits{J.}}:
\bbtitle{Mechanics of Motor Proteins and the Cytoskeleton}.
\bpublisher{Sinauer Associates Inc.},
\blocation{Sunderland, Mass.}
(\byear{2001})
\end{bbook}
\endbibitem

\bibitem{McMahon_Gallop_2005}
\begin{barticle}
\bauthor{\bsnm{McMahon}, \binits{H.T.}},
\bauthor{\bsnm{Gallop}, \binits{J.L.}}:
\batitle{Membrane curvature and mechanisms of dynamic cell membrane
  remodelling}.
\bjtitle{Nature}
\bvolume{438}(\bissue{7068}),
\bfpage{590}--\blpage{596}
(\byear{2005})
\end{barticle}
\endbibitem

\bibitem{Pollard_Mullins_2000}
\begin{barticle}
\bauthor{\bsnm{Pollard}, \binits{T.D.}},
\bauthor{\bsnm{Blanchoin}, \binits{L.}},
\bauthor{\bsnm{Mullins}, \binits{R.D.}}:
\batitle{Molecular mechanisms controlling actin filament dynamics in nonmuscle
  cells}.
\bjtitle{Annual review of biophysics and biomolecular structure}
\bvolume{29}(\bissue{1}),
\bfpage{545}--\blpage{576}
(\byear{2000})
\end{barticle}
\endbibitem

\bibitem{Dubyak_2004}
\begin{barticle}
\bauthor{\bsnm{Dubyak}, \binits{G.R.}}:
\batitle{Ion homeostasis, channels, and transporters: an update on cellular
  mechanisms}.
\bjtitle{Advances in physiology education}
\bvolume{28}(\bissue{4}),
\bfpage{143}--\blpage{154}
(\byear{2004})
\end{barticle}
\endbibitem

\bibitem{Wente_Rout_2010}
\begin{barticle}
\bauthor{\bsnm{Wente}, \binits{S.R.}},
\bauthor{\bsnm{Rout}, \binits{M.P.}}:
\batitle{The nuclear pore complex and nuclear transport}.
\bjtitle{Cold Spring Harbor perspectives in biology}
\bvolume{2}(\bissue{10}),
\bfpage{000562}
(\byear{2010})
\end{barticle}
\endbibitem

\bibitem{Hochstrasser_1995}
\begin{barticle}
\bauthor{\bsnm{Hochstrasser}, \binits{M.}}:
\batitle{Ubiquitin, proteasomes, and the regulation of intracellular protein
  degradation}.
\bjtitle{Current opinion in cell biology}
\bvolume{7}(\bissue{2}),
\bfpage{215}--\blpage{223}
(\byear{1995})
\end{barticle}
\endbibitem

\bibitem{Gu_Greenberg_2023}
\begin{barticle}
\bauthor{\bsnm{Gu}, \binits{X.}},
\bauthor{\bsnm{Nardone}, \binits{C.}},
\bauthor{\bsnm{Kamitaki}, \binits{N.}},
\bauthor{\bsnm{Mao}, \binits{A.}},
\bauthor{\bsnm{Elledge}, \binits{S.J.}},
\bauthor{\bsnm{Greenberg}, \binits{M.E.}}:
\batitle{The midnolin-proteasome pathway catches proteins for
  ubiquitination-independent degradation}.
\bjtitle{Science}
\bvolume{381}(\bissue{6660}),
\bfpage{5021}
(\byear{2023})
\end{barticle}
\endbibitem

\bibitem{Klionsky_Emr_2000}
\begin{barticle}
\bauthor{\bsnm{Klionsky}, \binits{D.J.}},
\bauthor{\bsnm{Emr}, \binits{S.D.}}:
\batitle{Autophagy as a regulated pathway of cellular degradation}.
\bjtitle{Science}
\bvolume{290}(\bissue{5497}),
\bfpage{1717}--\blpage{1721}
(\byear{2000})
\end{barticle}
\endbibitem

\bibitem{Nagata_2005}
\begin{barticle}
\bauthor{\bsnm{Nagata}, \binits{S.}}:
\batitle{Dna degradation in development and programmed cell death}.
\bjtitle{Annu. Rev. Immunol.}
\bvolume{23},
\bfpage{853}--\blpage{875}
(\byear{2005})
\end{barticle}
\endbibitem

\bibitem{Liu_Czaja_2013}
\begin{barticle}
\bauthor{\bsnm{Liu}, \binits{K.}},
\bauthor{\bsnm{Czaja}, \binits{M.}}:
\batitle{Regulation of lipid stores and metabolism by lipophagy}.
\bjtitle{Cell Death \& Differentiation}
\bvolume{20}(\bissue{1}),
\bfpage{3}--\blpage{11}
(\byear{2013})
\end{barticle}
\endbibitem

\bibitem{Neurath_Walsh_1976}
\begin{barticle}
\bauthor{\bsnm{Neurath}, \binits{H.}},
\bauthor{\bsnm{Walsh}, \binits{K.A.}}:
\batitle{Role of proteolytic enzymes in biological regulation (a review).}
\bjtitle{Proceedings of the National Academy of Sciences}
\bvolume{73}(\bissue{11}),
\bfpage{3825}--\blpage{3832}
(\byear{1976})
\end{barticle}
\endbibitem

\bibitem{rouille1995proteolytic}
\begin{barticle}
\bauthor{\bsnm{Rouill{\'e}}, \binits{Y.}},
\bauthor{\bsnm{Duguay}, \binits{S.J.}},
\bauthor{\bsnm{Lund}, \binits{K.}},
\bauthor{\bsnm{Furuta}, \binits{M.}},
\bauthor{\bsnm{Gong}, \binits{Q.}},
\bauthor{\bsnm{Lipkind}, \binits{G.}},
\bauthor{\bsnm{Oliva~Jr}, \binits{A.A.}},
\bauthor{\bsnm{Chan}, \binits{S.J.}},
\bauthor{\bsnm{Steiner}, \binits{D.F.}}:
\batitle{Proteolytic processing mechanisms in the biosynthesis of
  neuroendocrine peptides: the subtilisin-like proprotein convertases}.
\bjtitle{Frontiers in neuroendocrinology}
\bvolume{16}(\bissue{4}),
\bfpage{322}--\blpage{361}
(\byear{1995})
\end{barticle}
\endbibitem

\bibitem{singh2005autocrine}
\begin{barticle}
\bauthor{\bsnm{Singh}, \binits{A.B.}},
\bauthor{\bsnm{Harris}, \binits{R.C.}}:
\batitle{Autocrine, paracrine and juxtacrine signaling by egfr ligands}.
\bjtitle{Cellular signalling}
\bvolume{17}(\bissue{10}),
\bfpage{1183}--\blpage{1193}
(\byear{2005})
\end{barticle}
\endbibitem

\bibitem{Bolouri_Davidson_2002}
\begin{barticle}
\bauthor{\bsnm{Bolouri}, \binits{H.}},
\bauthor{\bsnm{Davidson}, \binits{E.H.}}:
\batitle{Modeling transcriptional regulatory networks}.
\bjtitle{BioEssays}
\bvolume{24}(\bissue{12}),
\bfpage{1118}--\blpage{1129}
(\byear{2002})
\end{barticle}
\endbibitem

\bibitem{Karlebach_Shamir_2008}
\begin{barticle}
\bauthor{\bsnm{Karlebach}, \binits{G.}},
\bauthor{\bsnm{Shamir}, \binits{R.}}:
\batitle{Modelling and analysis of gene regulatory networks}.
\bjtitle{Nature reviews Molecular cell biology}
\bvolume{9}(\bissue{10}),
\bfpage{770}--\blpage{780}
(\byear{2008})
\end{barticle}
\endbibitem

\bibitem{Trinh_Srienc_2008}
\begin{barticle}
\bauthor{\bsnm{Trinh}, \binits{C.T.}},
\bauthor{\bsnm{Unrean}, \binits{P.}},
\bauthor{\bsnm{Srienc}, \binits{F.}}:
\batitle{Minimal escherichia coli cell for the most efficient production of
  ethanol from hexoses and pentoses}.
\bjtitle{Applied and environmental microbiology}
\bvolume{74}(\bissue{12}),
\bfpage{3634}--\blpage{3643}
(\byear{2008})
\end{barticle}
\endbibitem

\bibitem{Kurjan_1992}
\begin{barticle}
\bauthor{\bsnm{Kurjan}, \binits{J.}}:
\batitle{Pheromone response in yeast}.
\bjtitle{Annual review of biochemistry}
\bvolume{61}(\bissue{1}),
\bfpage{1097}--\blpage{1129}
(\byear{1992})
\end{barticle}
\endbibitem

\bibitem{Wang_Dohlman_2004}
\begin{barticle}
\bauthor{\bsnm{Wang}, \binits{Y.}},
\bauthor{\bsnm{Dohlman}, \binits{H.G.}}:
\batitle{Pheromone signaling mechanisms in yeast: a prototypical sex machine}.
\bjtitle{Science}
\bvolume{306}(\bissue{5701}),
\bfpage{1508}--\blpage{1509}
(\byear{2004})
\end{barticle}
\endbibitem

\bibitem{Bardwell_2005}
\begin{barticle}
\bauthor{\bsnm{Bardwell}, \binits{L.}}:
\batitle{A walk-through of the yeast mating pheromone response pathway}.
\bjtitle{Peptides}
\bvolume{26}(\bissue{2}),
\bfpage{339}--\blpage{350}
(\byear{2005})
\end{barticle}
\endbibitem

\bibitem{Andrews_Arkin_2007}
\begin{barticle}
\bauthor{\bsnm{Andrews}, \binits{S.S.}},
\bauthor{\bsnm{Arkin}, \binits{A.P.}}:
\batitle{Systems biology: a switch for sex}.
\bjtitle{Current biology}
\bvolume{17}(\bissue{11}),
\bfpage{410}--\blpage{412}
(\byear{2007})
\end{barticle}
\endbibitem

\bibitem{Bush_Colman-Lerner_2016}
\begin{barticle}
\bauthor{\bsnm{Bush}, \binits{A.}},
\bauthor{\bsnm{Vasen}, \binits{G.}},
\bauthor{\bsnm{Constantinou}, \binits{A.}},
\bauthor{\bsnm{Dunayevich}, \binits{P.}},
\bauthor{\bsnm{Patop}, \binits{I.L.}},
\bauthor{\bsnm{Blaustein}, \binits{M.}},
\bauthor{\bsnm{Colman-Lerner}, \binits{A.}}:
\batitle{Yeast gpcr signaling reflects the fraction of occupied receptors, not
  the number}.
\bjtitle{Molecular systems biology}
\bvolume{12}(\bissue{12}),
\bfpage{898}
(\byear{2016})
\end{barticle}
\endbibitem

\bibitem{Chan_Kyba_2013}
\begin{botherref}
\oauthor{\bsnm{Chan}, \binits{S.S.-K.}},
\oauthor{\bsnm{Kyba}, \binits{M.}}:
What is a master regulator?
Journal of stem cell research \& therapy
\textbf{3}
(2013)
\end{botherref}
\endbibitem

\bibitem{lichtenthaler2018proteolytic}
\begin{barticle}
\bauthor{\bsnm{Lichtenthaler}, \binits{S.F.}},
\bauthor{\bsnm{Lemberg}, \binits{M.K.}},
\bauthor{\bsnm{Fluhrer}, \binits{R.}}:
\batitle{Proteolytic ectodomain shedding of membrane proteins in
  mammals-hardware, concepts, and recent developments}.
\bjtitle{The EMBO journal}
\bvolume{37}(\bissue{15}),
\bfpage{99456}
(\byear{2018})
\end{barticle}
\endbibitem

\bibitem{Loewith_Hall_2011}
\begin{barticle}
\bauthor{\bsnm{Loewith}, \binits{R.}},
\bauthor{\bsnm{Hall}, \binits{M.N.}}:
\batitle{Target of rapamycin (tor) in nutrient signaling and growth control}.
\bjtitle{Genetics}
\bvolume{189}(\bissue{4}),
\bfpage{1177}--\blpage{1201}
(\byear{2011})
\end{barticle}
\endbibitem

\bibitem{Duke_Bray_2001}
\begin{barticle}
\bauthor{\bsnm{Duke}, \binits{T.}},
\bauthor{\bsnm{Le~Novere}, \binits{N.}},
\bauthor{\bsnm{Bray}, \binits{D.}}:
\batitle{Conformational spread in a ring of proteins: a stochastic approach to
  allostery}.
\bjtitle{Journal of molecular biology}
\bvolume{308}(\bissue{3}),
\bfpage{541}--\blpage{553}
(\byear{2001})
\end{barticle}
\endbibitem

\bibitem{shi2016conservation}
\begin{barticle}
\bauthor{\bsnm{Shi}, \binits{T.}},
\bauthor{\bsnm{Niepel}, \binits{M.}},
\bauthor{\bsnm{McDermott}, \binits{J.E.}},
\bauthor{\bsnm{Gao}, \binits{Y.}},
\bauthor{\bsnm{Nicora}, \binits{C.D.}},
\bauthor{\bsnm{Chrisler}, \binits{W.B.}},
\bauthor{\bsnm{Markillie}, \binits{L.M.}},
\bauthor{\bsnm{Petyuk}, \binits{V.A.}},
\bauthor{\bsnm{Smith}, \binits{R.D.}},
\bauthor{\bsnm{Rodland}, \binits{K.D.}}, \betal:
\batitle{Conservation of protein abundance patterns reveals the regulatory
  architecture of the egfr-mapk pathway}.
\bjtitle{Science Signaling}
\bvolume{9}(\bissue{436}),
\bfpage{6}--\blpage{6}
(\byear{2016})
\end{barticle}
\endbibitem

\bibitem{Csete_Doyle_2004}
\begin{barticle}
\bauthor{\bsnm{Csete}, \binits{M.}},
\bauthor{\bsnm{Doyle}, \binits{J.}}:
\batitle{Bow ties, metabolism and disease}.
\bjtitle{TRENDS in Biotechnology}
\bvolume{22}(\bissue{9}),
\bfpage{446}--\blpage{450}
(\byear{2004})
\end{barticle}
\endbibitem

\bibitem{Goentoro_Alon_2009}
\begin{barticle}
\bauthor{\bsnm{Goentoro}, \binits{L.}},
\bauthor{\bsnm{Shoval}, \binits{O.}},
\bauthor{\bsnm{Kirschner}, \binits{M.W.}},
\bauthor{\bsnm{Alon}, \binits{U.}}:
\batitle{The incoherent feedforward loop can provide fold-change detection in
  gene regulation}.
\bjtitle{Molecular cell}
\bvolume{36}(\bissue{5}),
\bfpage{894}--\blpage{899}
(\byear{2009})
\end{barticle}
\endbibitem

\bibitem{Mangan_Alon_2003}
\begin{barticle}
\bauthor{\bsnm{Mangan}, \binits{S.}},
\bauthor{\bsnm{Zaslaver}, \binits{A.}},
\bauthor{\bsnm{Alon}, \binits{U.}}:
\batitle{The coherent feedforward loop serves as a sign-sensitive delay element
  in transcription networks}.
\bjtitle{Journal of molecular biology}
\bvolume{334}(\bissue{2}),
\bfpage{197}--\blpage{204}
(\byear{2003})
\end{barticle}
\endbibitem

\bibitem{Andrews_Brent_2016}
\begin{barticle}
\bauthor{\bsnm{Andrews}, \binits{S.S.}},
\bauthor{\bsnm{Peria}, \binits{W.J.}},
\bauthor{\bsnm{Richard}, \binits{C.Y.}},
\bauthor{\bsnm{Colman-Lerner}, \binits{A.}},
\bauthor{\bsnm{Brent}, \binits{R.}}:
\batitle{Push-pull and feedback mechanisms can align signaling system outputs
  with inputs}.
\bjtitle{Cell systems}
\bvolume{3}(\bissue{5}),
\bfpage{444}--\blpage{455}
(\byear{2016})
\end{barticle}
\endbibitem

\bibitem{Sauro_2017}
\begin{barticle}
\bauthor{\bsnm{Sauro}, \binits{H.M.}}:
\batitle{Control and regulation of pathways via negative feedback}.
\bjtitle{Journal of The Royal Society Interface}
\bvolume{14}(\bissue{127}),
\bfpage{20160848}
(\byear{2017})
\end{barticle}
\endbibitem

\bibitem{Dohlman_Thorner_1996}
\begin{barticle}
\bauthor{\bsnm{Dohlman}, \binits{H.G.}},
\bauthor{\bsnm{Song}, \binits{J.}},
\bauthor{\bsnm{Ma}, \binits{D.}},
\bauthor{\bsnm{Courchesne}, \binits{W.E.}},
\bauthor{\bsnm{Thorner}, \binits{J.}}:
\batitle{Sst2, a negative regulator of pheromone signaling in the yeast
  saccharomyces cerevisiae: expression, localization, and genetic interaction
  and physical association with gpa1 (the g-protein alpha subunit)}.
\bjtitle{Molecular and cellular biology}
\bvolume{16}(\bissue{9}),
\bfpage{5194}--\blpage{5209}
(\byear{1996})
\end{barticle}
\endbibitem

\bibitem{Venkatapurapu_Elston_2015}
\begin{barticle}
\bauthor{\bsnm{Venkatapurapu}, \binits{S.P.}},
\bauthor{\bsnm{Kelley}, \binits{J.B.}},
\bauthor{\bsnm{Dixit}, \binits{G.}},
\bauthor{\bsnm{Pena}, \binits{M.}},
\bauthor{\bsnm{Errede}, \binits{B.}},
\bauthor{\bsnm{Dohlman}, \binits{H.G.}},
\bauthor{\bsnm{Elston}, \binits{T.C.}}:
\batitle{Modulation of receptor dynamics by the regulator of g protein
  signaling sst2}.
\bjtitle{Molecular biology of the cell}
\bvolume{26}(\bissue{22}),
\bfpage{4124}--\blpage{4134}
(\byear{2015})
\end{barticle}
\endbibitem

\bibitem{Dixit_Dohlman_2014}
\begin{barticle}
\bauthor{\bsnm{Dixit}, \binits{G.}},
\bauthor{\bsnm{Kelley}, \binits{J.B.}},
\bauthor{\bsnm{Houser}, \binits{J.R.}},
\bauthor{\bsnm{Elston}, \binits{T.C.}},
\bauthor{\bsnm{Dohlman}, \binits{H.G.}}:
\batitle{Cellular noise suppression by the regulator of g protein signaling
  sst2}.
\bjtitle{Molecular cell}
\bvolume{55}(\bissue{1}),
\bfpage{85}--\blpage{96}
(\byear{2014})
\end{barticle}
\endbibitem

\bibitem{Sourjik_Berg_2002}
\begin{barticle}
\bauthor{\bsnm{Sourjik}, \binits{V.}},
\bauthor{\bsnm{Berg}, \binits{H.C.}}:
\batitle{Receptor sensitivity in bacterial chemotaxis}.
\bjtitle{Proceedings of the National Academy of Sciences}
\bvolume{99}(\bissue{1}),
\bfpage{123}--\blpage{127}
(\byear{2002})
\end{barticle}
\endbibitem

\bibitem{Paik_Kim_2007}
\begin{barticle}
\bauthor{\bsnm{Paik}, \binits{W.K.}},
\bauthor{\bsnm{Paik}, \binits{D.C.}},
\bauthor{\bsnm{Kim}, \binits{S.}}:
\batitle{Historical review: the field of protein methylation}.
\bjtitle{Trends in biochemical sciences}
\bvolume{32}(\bissue{3}),
\bfpage{146}--\blpage{152}
(\byear{2007})
\end{barticle}
\endbibitem

\bibitem{Moore_Fan_2013}
\begin{barticle}
\bauthor{\bsnm{Moore}, \binits{L.D.}},
\bauthor{\bsnm{Le}, \binits{T.}},
\bauthor{\bsnm{Fan}, \binits{G.}}:
\batitle{Dna methylation and its basic function}.
\bjtitle{Neuropsychopharmacology}
\bvolume{38}(\bissue{1}),
\bfpage{23}--\blpage{38}
(\byear{2013})
\end{barticle}
\endbibitem

\bibitem{Sloan_Bohnsack_2017}
\begin{barticle}
\bauthor{\bsnm{Sloan}, \binits{K.E.}},
\bauthor{\bsnm{Warda}, \binits{A.S.}},
\bauthor{\bsnm{Sharma}, \binits{S.}},
\bauthor{\bsnm{Entian}, \binits{K.-D.}},
\bauthor{\bsnm{Lafontaine}, \binits{D.L.}},
\bauthor{\bsnm{Bohnsack}, \binits{M.T.}}:
\batitle{Tuning the ribosome: the influence of rrna modification on eukaryotic
  ribosome biogenesis and function}.
\bjtitle{RNA biology}
\bvolume{14}(\bissue{9}),
\bfpage{1138}--\blpage{1152}
(\byear{2017})
\end{barticle}
\endbibitem

\bibitem{Lutkenhaus_2007}
\begin{barticle}
\bauthor{\bsnm{Lutkenhaus}, \binits{J.}}:
\batitle{Assembly dynamics of the bacterial mincde system and spatial
  regulation of the z ring}.
\bjtitle{Annu. Rev. Biochem.}
\bvolume{76},
\bfpage{539}--\blpage{562}
(\byear{2007})
\end{barticle}
\endbibitem

\bibitem{Groban_Jacobson_2006}
\begin{barticle}
\bauthor{\bsnm{Groban}, \binits{E.S.}},
\bauthor{\bsnm{Narayanan}, \binits{A.}},
\bauthor{\bsnm{Jacobson}, \binits{M.P.}}:
\batitle{Conformational changes in protein loops and helices induced by
  post-translational phosphorylation}.
\bjtitle{PLoS computational biology}
\bvolume{2}(\bissue{4}),
\bfpage{32}
(\byear{2006})
\end{barticle}
\endbibitem

\bibitem{Huang_Ferrell_1996}
\begin{barticle}
\bauthor{\bsnm{Huang}, \binits{C.-Y.}},
\bauthor{\bsnm{Ferrell~Jr}, \binits{J.E.}}:
\batitle{Ultrasensitivity in the mitogen-activated protein kinase cascade.}
\bjtitle{Proceedings of the National Academy of Sciences}
\bvolume{93}(\bissue{19}),
\bfpage{10078}--\blpage{10083}
(\byear{1996})
\end{barticle}
\endbibitem

\bibitem{Roach_1991}
\begin{barticle}
\bauthor{\bsnm{Roach}, \binits{P.}}:
\batitle{Multisite and hierarchal protein phosphorylation.}
\bjtitle{The Journal of biological chemistry}
\bvolume{266}(\bissue{22}),
\bfpage{14139}--\blpage{14142}
(\byear{1991})
\end{barticle}
\endbibitem

\bibitem{Mann_Pandey_2002}
\begin{barticle}
\bauthor{\bsnm{Mann}, \binits{M.}},
\bauthor{\bsnm{Ong}, \binits{S.-E.}},
\bauthor{\bsnm{Gr{\o}nborg}, \binits{M.}},
\bauthor{\bsnm{Steen}, \binits{H.}},
\bauthor{\bsnm{Jensen}, \binits{O.N.}},
\bauthor{\bsnm{Pandey}, \binits{A.}}:
\batitle{Analysis of protein phosphorylation using mass spectrometry:
  deciphering the phosphoproteome}.
\bjtitle{Trends in biotechnology}
\bvolume{20}(\bissue{6}),
\bfpage{261}--\blpage{268}
(\byear{2002})
\end{barticle}
\endbibitem

\bibitem{Kapuy_Novak_2009}
\begin{barticle}
\bauthor{\bsnm{Kapuy}, \binits{O.}},
\bauthor{\bsnm{Barik}, \binits{D.}},
\bauthor{\bsnm{Sananes}, \binits{M.R.D.}},
\bauthor{\bsnm{Tyson}, \binits{J.J.}},
\bauthor{\bsnm{Nov{\'a}k}, \binits{B.}}:
\batitle{Bistability by multiple phosphorylation of regulatory proteins}.
\bjtitle{Progress in biophysics and molecular biology}
\bvolume{100}(\bissue{1-3}),
\bfpage{47}--\blpage{56}
(\byear{2009})
\end{barticle}
\endbibitem

\bibitem{McPherson_Turnbough_2010}
\begin{barticle}
\bauthor{\bsnm{McPherson}, \binits{S.A.}},
\bauthor{\bsnm{Li}, \binits{M.}},
\bauthor{\bsnm{Kearney}, \binits{J.F.}},
\bauthor{\bsnm{Turnbough~Jr}, \binits{C.L.}}:
\batitle{Exsb, an unusually highly phosphorylated protein required for the
  stable attachment of the exosporium of bacillus anthracis}.
\bjtitle{Molecular microbiology}
\bvolume{76}(\bissue{6}),
\bfpage{1527}--\blpage{1538}
(\byear{2010})
\end{barticle}
\endbibitem

\bibitem{Rosen_Ron_2004}
\begin{barticle}
\bauthor{\bsnm{Rosen}, \binits{R.}},
\bauthor{\bsnm{Becher}, \binits{D.}},
\bauthor{\bsnm{B{\"u}ttner}, \binits{K.}},
\bauthor{\bsnm{Biran}, \binits{D.}},
\bauthor{\bsnm{Hecker}, \binits{M.}},
\bauthor{\bsnm{Ron}, \binits{E.Z.}}:
\batitle{Highly phosphorylated bacterial proteins}.
\bjtitle{Proteomics}
\bvolume{4}(\bissue{10}),
\bfpage{3068}--\blpage{3077}
(\byear{2004})
\end{barticle}
\endbibitem

\bibitem{vanderGeer_Pawson_1996}
\begin{barticle}
\bauthor{\bparticle{van~der} \bsnm{Geer}, \binits{P.}},
\bauthor{\bsnm{Wiley}, \binits{S.}},
\bauthor{\bsnm{Gish}, \binits{G.D.}},
\bauthor{\bsnm{Pawson}, \binits{T.}}:
\batitle{The shc adaptor protein is highly phosphorylated at conserved, twin
  tyrosine residues (y239/240) that mediate protein--protein interactions}.
\bjtitle{Current Biology}
\bvolume{6}(\bissue{11}),
\bfpage{1435}--\blpage{1444}
(\byear{1996})
\end{barticle}
\endbibitem

\bibitem{Moraru_Loew_2005}
\begin{barticle}
\bauthor{\bsnm{Moraru}, \binits{I.I.}},
\bauthor{\bsnm{Loew}, \binits{L.M.}}:
\batitle{Intracellular signaling: spatial and temporal control}.
\bjtitle{Physiology}
\bvolume{20}(\bissue{3}),
\bfpage{169}--\blpage{179}
(\byear{2005})
\end{barticle}
\endbibitem

\bibitem{Dehmelt_Bastiaens_2010}
\begin{barticle}
\bauthor{\bsnm{Dehmelt}, \binits{L.}},
\bauthor{\bsnm{Bastiaens}, \binits{P.I.}}:
\batitle{Spatial organization of intracellular communication: insights from
  imaging}.
\bjtitle{Nature reviews Molecular cell biology}
\bvolume{11}(\bissue{6}),
\bfpage{440}--\blpage{452}
(\byear{2010})
\end{barticle}
\endbibitem

\bibitem{Rao_Arkin_2002}
\begin{barticle}
\bauthor{\bsnm{Rao}, \binits{C.V.}},
\bauthor{\bsnm{Wolf}, \binits{D.M.}},
\bauthor{\bsnm{Arkin}, \binits{A.P.}}:
\batitle{Control, exploitation and tolerance of intracellular noise}.
\bjtitle{Nature}
\bvolume{420}(\bissue{6912}),
\bfpage{231}--\blpage{237}
(\byear{2002})
\end{barticle}
\endbibitem

\bibitem{Elowitz_Swain_2002}
\begin{barticle}
\bauthor{\bsnm{Elowitz}, \binits{M.B.}},
\bauthor{\bsnm{Levine}, \binits{A.J.}},
\bauthor{\bsnm{Siggia}, \binits{E.D.}},
\bauthor{\bsnm{Swain}, \binits{P.S.}}:
\batitle{Stochastic gene expression in a single cell}.
\bjtitle{Science}
\bvolume{297}(\bissue{5584}),
\bfpage{1183}--\blpage{1186}
(\byear{2002})
\end{barticle}
\endbibitem

\bibitem{Tsimring_2014}
\begin{barticle}
\bauthor{\bsnm{Tsimring}, \binits{L.S.}}:
\batitle{Noise in biology}.
\bjtitle{Reports on Progress in Physics}
\bvolume{77}(\bissue{2}),
\bfpage{026601}
(\byear{2014})
\end{barticle}
\endbibitem

\bibitem{Muzzey_vanOudenaarden_2009}
\begin{barticle}
\bauthor{\bsnm{Muzzey}, \binits{D.}},
\bauthor{\bsnm{G{\'o}mez-Uribe}, \binits{C.A.}},
\bauthor{\bsnm{Mettetal}, \binits{J.T.}},
\bauthor{\bparticle{van} \bsnm{Oudenaarden}, \binits{A.}}:
\batitle{A systems-level analysis of perfect adaptation in yeast
  osmoregulation}.
\bjtitle{Cell}
\bvolume{138}(\bissue{1}),
\bfpage{160}--\blpage{171}
(\byear{2009})
\end{barticle}
\endbibitem

\bibitem{Saunders_Wessels_1998}
\begin{barticle}
\bauthor{\bsnm{Saunders}, \binits{P.T.}},
\bauthor{\bsnm{Koeslag}, \binits{J.H.}},
\bauthor{\bsnm{Wessels}, \binits{J.A.}}:
\batitle{Integral rein control in physiology}.
\bjtitle{Journal of Theoretical Biology}
\bvolume{194}(\bissue{2}),
\bfpage{163}--\blpage{173}
(\byear{1998})
\end{barticle}
\endbibitem

\bibitem{Huang_Ruoff_2012}
\begin{barticle}
\bauthor{\bsnm{Huang}, \binits{Y.}},
\bauthor{\bsnm{Drengstig}, \binits{T.}},
\bauthor{\bsnm{Ruoff}, \binits{P.}}:
\batitle{Integrating fluctuating nitrate uptake and assimilation to robust
  homeostasis}.
\bjtitle{Plant, Cell \& Environment}
\bvolume{35}(\bissue{5}),
\bfpage{917}--\blpage{928}
(\byear{2012})
\end{barticle}
\endbibitem

\bibitem{Jin_Elston_2011}
\begin{barticle}
\bauthor{\bsnm{Jin}, \binits{M.}},
\bauthor{\bsnm{Errede}, \binits{B.}},
\bauthor{\bsnm{Behar}, \binits{M.}},
\bauthor{\bsnm{Mather}, \binits{W.}},
\bauthor{\bsnm{Nayak}, \binits{S.}},
\bauthor{\bsnm{Hasty}, \binits{J.}},
\bauthor{\bsnm{Dohlman}, \binits{H.G.}},
\bauthor{\bsnm{Elston}, \binits{T.C.}}:
\batitle{Yeast dynamically modify their environment to achieve better mating
  efficiency}.
\bjtitle{Science signaling}
\bvolume{4}(\bissue{186}),
\bfpage{54}--\blpage{54}
(\byear{2011})
\end{barticle}
\endbibitem

\bibitem{Yi_Doyle_2000}
\begin{barticle}
\bauthor{\bsnm{Yi}, \binits{T.-M.}},
\bauthor{\bsnm{Huang}, \binits{Y.}},
\bauthor{\bsnm{Simon}, \binits{M.I.}},
\bauthor{\bsnm{Doyle}, \binits{J.}}:
\batitle{Robust perfect adaptation in bacterial chemotaxis through integral
  feedback control}.
\bjtitle{Proceedings of the National Academy of Sciences}
\bvolume{97}(\bissue{9}),
\bfpage{4649}--\blpage{4653}
(\byear{2000})
\end{barticle}
\endbibitem

\bibitem{Qian_DelVecchio_2018}
\begin{barticle}
\bauthor{\bsnm{Qian}, \binits{Y.}},
\bauthor{\bsnm{Del~Vecchio}, \binits{D.}}:
\batitle{Realizing `integral control' in living cells: how to overcome leaky
  integration due to dilution?}
\bjtitle{Journal of The Royal Society Interface}
\bvolume{15}(\bissue{139}),
\bfpage{20170902}
(\byear{2018})
\end{barticle}
\endbibitem

\bibitem{Briat_Khammash_2016}
\begin{barticle}
\bauthor{\bsnm{Briat}, \binits{C.}},
\bauthor{\bsnm{Gupta}, \binits{A.}},
\bauthor{\bsnm{Khammash}, \binits{M.}}:
\batitle{Antithetic integral feedback ensures robust perfect adaptation in
  noisy biomolecular networks}.
\bjtitle{Cell systems}
\bvolume{2}(\bissue{1}),
\bfpage{15}--\blpage{26}
(\byear{2016})
\end{barticle}
\endbibitem

\bibitem{Tu_McKnight_2006}
\begin{barticle}
\bauthor{\bsnm{Tu}, \binits{B.P.}},
\bauthor{\bsnm{McKnight}, \binits{S.L.}}:
\batitle{Metabolic cycles as an underlying basis of biological oscillations}.
\bjtitle{Nature reviews Molecular cell biology}
\bvolume{7}(\bissue{9}),
\bfpage{696}--\blpage{701}
(\byear{2006})
\end{barticle}
\endbibitem

\bibitem{Nelson_White_2004}
\begin{barticle}
\bauthor{\bsnm{Nelson}, \binits{D.}},
\bauthor{\bsnm{Ihekwaba}, \binits{A.}},
\bauthor{\bsnm{Elliott}, \binits{M.}},
\bauthor{\bsnm{Johnson}, \binits{J.}},
\bauthor{\bsnm{Gibney}, \binits{C.}},
\bauthor{\bsnm{Foreman}, \binits{B.}},
\bauthor{\bsnm{Nelson}, \binits{G.}},
\bauthor{\bsnm{See}, \binits{V.}},
\bauthor{\bsnm{Horton}, \binits{C.}},
\bauthor{\bsnm{Spiller}, \binits{D.}}, \betal:
\batitle{Oscillations in nf-$\kappa$b signaling control the dynamics of gene
  expression}.
\bjtitle{Science}
\bvolume{306}(\bissue{5696}),
\bfpage{704}--\blpage{708}
(\byear{2004})
\end{barticle}
\endbibitem

\bibitem{Shankaran_Wiley_2009}
\begin{barticle}
\bauthor{\bsnm{Shankaran}, \binits{H.}},
\bauthor{\bsnm{Ippolito}, \binits{D.L.}},
\bauthor{\bsnm{Chrisler}, \binits{W.B.}},
\bauthor{\bsnm{Resat}, \binits{H.}},
\bauthor{\bsnm{Bollinger}, \binits{N.}},
\bauthor{\bsnm{Opresko}, \binits{L.K.}},
\bauthor{\bsnm{Wiley}, \binits{H.S.}}:
\batitle{Rapid and sustained nuclear--cytoplasmic erk oscillations induced by
  epidermal growth factor}.
\bjtitle{Molecular systems biology}
\bvolume{5}(\bissue{1}),
\bfpage{332}
(\byear{2009})
\end{barticle}
\endbibitem

\bibitem{Isomura_Kageyama_2014}
\begin{barticle}
\bauthor{\bsnm{Isomura}, \binits{A.}},
\bauthor{\bsnm{Kageyama}, \binits{R.}}:
\batitle{Ultradian oscillations and pulses: coordinating cellular responses and
  cell fate decisions}.
\bjtitle{Development}
\bvolume{141}(\bissue{19}),
\bfpage{3627}--\blpage{3636}
(\byear{2014})
\end{barticle}
\endbibitem

\bibitem{Leloup_Goldbeter_2003}
\begin{barticle}
\bauthor{\bsnm{Leloup}, \binits{J.-C.}},
\bauthor{\bsnm{Goldbeter}, \binits{A.}}:
\batitle{Toward a detailed computational model for the mammalian circadian
  clock}.
\bjtitle{Proceedings of the National Academy of Sciences}
\bvolume{100}(\bissue{12}),
\bfpage{7051}--\blpage{7056}
(\byear{2003})
\end{barticle}
\endbibitem

\bibitem{Elowitz_Leibler_2000}
\begin{barticle}
\bauthor{\bsnm{Elowitz}, \binits{M.B.}},
\bauthor{\bsnm{Leibler}, \binits{S.}}:
\batitle{A synthetic oscillatory network of transcriptional regulators}.
\bjtitle{Nature}
\bvolume{403}(\bissue{6767}),
\bfpage{335}--\blpage{338}
(\byear{2000})
\end{barticle}
\endbibitem

\bibitem{Chickarmane_Sauro_2007}
\begin{barticle}
\bauthor{\bsnm{Chickarmane}, \binits{V.}},
\bauthor{\bsnm{Kholodenko}, \binits{B.N.}},
\bauthor{\bsnm{Sauro}, \binits{H.M.}}:
\batitle{Oscillatory dynamics arising from competitive inhibition and multisite
  phosphorylation}.
\bjtitle{Journal of theoretical biology}
\bvolume{244}(\bissue{1}),
\bfpage{68}--\blpage{76}
(\byear{2007})
\end{barticle}
\endbibitem

\bibitem{Novak_Tyson_2008}
\begin{barticle}
\bauthor{\bsnm{Nov{\'a}k}, \binits{B.}},
\bauthor{\bsnm{Tyson}, \binits{J.J.}}:
\batitle{Design principles of biochemical oscillators}.
\bjtitle{Nature reviews Molecular cell biology}
\bvolume{9}(\bissue{12}),
\bfpage{981}--\blpage{991}
(\byear{2008})
\end{barticle}
\endbibitem

\bibitem{Tyson_1991}
\begin{barticle}
\bauthor{\bsnm{Tyson}, \binits{J.J.}}:
\batitle{Modeling the cell division cycle: cdc2 and cyclin interactions.}
\bjtitle{Proceedings of the National Academy of Sciences}
\bvolume{88}(\bissue{16}),
\bfpage{7328}--\blpage{7332}
(\byear{1991})
\end{barticle}
\endbibitem

\bibitem{Goldbeter_Berridge_1990}
\begin{barticle}
\bauthor{\bsnm{Goldbeter}, \binits{A.}},
\bauthor{\bsnm{Dupont}, \binits{G.}},
\bauthor{\bsnm{Berridge}, \binits{M.J.}}:
\batitle{Minimal model for signal-induced ca2+ oscillations and for their
  frequency encoding through protein phosphorylation.}
\bjtitle{Proceedings of the National Academy of Sciences}
\bvolume{87}(\bissue{4}),
\bfpage{1461}--\blpage{1465}
(\byear{1990})
\end{barticle}
\endbibitem

\bibitem{Halloy_Goldbeter_1998}
\begin{barticle}
\bauthor{\bsnm{Halloy}, \binits{J.}},
\bauthor{\bsnm{Lauzeral}, \binits{J.}},
\bauthor{\bsnm{Goldbeter}, \binits{A.}}:
\batitle{Modeling oscillations and waves of camp in dictyostelium discoideum
  cells}.
\bjtitle{Biophysical chemistry}
\bvolume{72}(\bissue{1-2}),
\bfpage{9}--\blpage{19}
(\byear{1998})
\end{barticle}
\endbibitem

\bibitem{Levine_Elowitz_2013}
\begin{barticle}
\bauthor{\bsnm{Levine}, \binits{J.H.}},
\bauthor{\bsnm{Lin}, \binits{Y.}},
\bauthor{\bsnm{Elowitz}, \binits{M.B.}}:
\batitle{Functional roles of pulsing in genetic circuits}.
\bjtitle{Science}
\bvolume{342}(\bissue{6163}),
\bfpage{1193}--\blpage{1200}
(\byear{2013})
\end{barticle}
\endbibitem

\bibitem{sauro2007mapk}
\begin{botherref}
\oauthor{\bsnm{Sauro}, \binits{H.M.}},
\oauthor{\bsnm{Ingalls}, \binits{B.}}:
{MAPK} cascades as feedback amplifiers.
arXiv preprint arXiv:0710.5195
(2007)
\end{botherref}
\endbibitem

\bibitem{Andrews_Balazsi_2018}
\begin{barticle}
\bauthor{\bsnm{Andrews}, \binits{S.S.}},
\bauthor{\bsnm{Brent}, \binits{R.}},
\bauthor{\bsnm{Bal{\'a}zsi}, \binits{G.}}:
\batitle{Transferring information without distortion}.
\bjtitle{Elife}
\bvolume{7},
\bfpage{41894}
(\byear{2018})
\end{barticle}
\endbibitem

\bibitem{Yi_Simon_2003}
\begin{barticle}
\bauthor{\bsnm{Yi}, \binits{T.-M.}},
\bauthor{\bsnm{Kitano}, \binits{H.}},
\bauthor{\bsnm{Simon}, \binits{M.I.}}:
\batitle{A quantitative characterization of the yeast heterotrimeric g protein
  cycle}.
\bjtitle{Proceedings of the National Academy of Sciences}
\bvolume{100}(\bissue{19}),
\bfpage{10764}--\blpage{10769}
(\byear{2003})
\end{barticle}
\endbibitem

\bibitem{Yu_Brent_2008}
\begin{barticle}
\bauthor{\bsnm{Yu}, \binits{R.C.}},
\bauthor{\bsnm{Pesce}, \binits{C.G.}},
\bauthor{\bsnm{Colman-Lerner}, \binits{A.}},
\bauthor{\bsnm{Lok}, \binits{L.}},
\bauthor{\bsnm{Pincus}, \binits{D.}},
\bauthor{\bsnm{Serra}, \binits{E.}},
\bauthor{\bsnm{Holl}, \binits{M.}},
\bauthor{\bsnm{Benjamin}, \binits{K.}},
\bauthor{\bsnm{Gordon}, \binits{A.}},
\bauthor{\bsnm{Brent}, \binits{R.}}:
\batitle{Negative feedback that improves information transmission in yeast
  signalling}.
\bjtitle{Nature}
\bvolume{456}(\bissue{7223}),
\bfpage{755}--\blpage{761}
(\byear{2008})
\end{barticle}
\endbibitem

\bibitem{Nunns_Goentoro_2018}
\begin{barticle}
\bauthor{\bsnm{Nunns}, \binits{H.}},
\bauthor{\bsnm{Goentoro}, \binits{L.}}:
\batitle{Signaling pathways as linear transmitters}.
\bjtitle{Elife}
\bvolume{7},
\bfpage{33617}
(\byear{2018})
\end{barticle}
\endbibitem

\bibitem{Black_Leff_1983}
\begin{barticle}
\bauthor{\bsnm{Black}, \binits{J.W.}},
\bauthor{\bsnm{Leff}, \binits{P.}}:
\batitle{Operational models of pharmacological agonism}.
\bjtitle{Proceedings of the Royal society of London. Series B. Biological
  sciences}
\bvolume{220}(\bissue{1219}),
\bfpage{141}--\blpage{162}
(\byear{1983})
\end{barticle}
\endbibitem

\bibitem{Nevozhay_Balazsi_2009}
\begin{barticle}
\bauthor{\bsnm{Nevozhay}, \binits{D.}},
\bauthor{\bsnm{Adams}, \binits{R.M.}},
\bauthor{\bsnm{Murphy}, \binits{K.F.}},
\bauthor{\bsnm{Josi{\'c}}, \binits{K.}},
\bauthor{\bsnm{Bal{\'a}zsi}, \binits{G.}}:
\batitle{Negative autoregulation linearizes the dose--response and suppresses
  the heterogeneity of gene expression}.
\bjtitle{Proceedings of the National Academy of Sciences}
\bvolume{106}(\bissue{13}),
\bfpage{5123}--\blpage{5128}
(\byear{2009})
\end{barticle}
\endbibitem

\bibitem{Conde_Ovsenek_2009}
\begin{barticle}
\bauthor{\bsnm{Conde}, \binits{R.}},
\bauthor{\bsnm{Belak}, \binits{Z.R.}},
\bauthor{\bsnm{Nair}, \binits{M.}},
\bauthor{\bsnm{O'Carroll}, \binits{R.F.}},
\bauthor{\bsnm{Ovsenek}, \binits{N.}}:
\batitle{Modulation of hsf1 activity by novobiocin and geldanamycin}.
\bjtitle{Biochemistry and Cell Biology}
\bvolume{87}(\bissue{6}),
\bfpage{845}--\blpage{851}
(\byear{2009})
\end{barticle}
\endbibitem

\bibitem{Bray_Morton-Firth_1998}
\begin{barticle}
\bauthor{\bsnm{Bray}, \binits{D.}},
\bauthor{\bsnm{Levin}, \binits{M.D.}},
\bauthor{\bsnm{Morton-Firth}, \binits{C.J.}}:
\batitle{Receptor clustering as a cellular mechanism to control sensitivity}.
\bjtitle{Nature}
\bvolume{393}(\bissue{6680}),
\bfpage{85}--\blpage{88}
(\byear{1998})
\end{barticle}
\endbibitem

\bibitem{small1990covalent}
\begin{barticle}
\bauthor{\bsnm{Small}, \binits{J.R.}},
\bauthor{\bsnm{Fell}, \binits{D.A.}}:
\batitle{Covalent modification and metabolic control analysis: Modification to
  the theorems and their application to metabolic systems containing covalently
  modifiable enzymes}.
\bjtitle{European journal of biochemistry}
\bvolume{191}(\bissue{2}),
\bfpage{405}--\blpage{411}
(\byear{1990})
\end{barticle}
\endbibitem

\bibitem{Zhang_Anderson_2013}
\begin{barticle}
\bauthor{\bsnm{Zhang}, \binits{Q.}},
\bauthor{\bsnm{Bhattacharya}, \binits{S.}},
\bauthor{\bsnm{Andersen}, \binits{M.E.}}:
\batitle{Ultrasensitive response motifs: basic amplifiers in molecular
  signalling networks}.
\bjtitle{Open biology}
\bvolume{3}(\bissue{4}),
\bfpage{130031}
(\byear{2013})
\end{barticle}
\endbibitem

\bibitem{Goutelle_Maire_2008}
\begin{barticle}
\bauthor{\bsnm{Goutelle}, \binits{S.}},
\bauthor{\bsnm{Maurin}, \binits{M.}},
\bauthor{\bsnm{Rougier}, \binits{F.}},
\bauthor{\bsnm{Barbaut}, \binits{X.}},
\bauthor{\bsnm{Bourguignon}, \binits{L.}},
\bauthor{\bsnm{Ducher}, \binits{M.}},
\bauthor{\bsnm{Maire}, \binits{P.}}:
\batitle{The hill equation: a review of its capabilities in pharmacological
  modelling}.
\bjtitle{Fundamental \& clinical pharmacology}
\bvolume{22}(\bissue{6}),
\bfpage{633}--\blpage{648}
(\byear{2008})
\end{barticle}
\endbibitem

\bibitem{hofmeyr1997reversible}
\begin{barticle}
\bauthor{\bsnm{Hofmeyr}, \binits{J.-H.S.}},
\bauthor{\bsnm{Cornish-Bowden}, \binits{H.}}:
\batitle{The reversible hill equation: how to incorporate cooperative enzymes
  into metabolic models}.
\bjtitle{Bioinformatics}
\bvolume{13}(\bissue{4}),
\bfpage{377}--\blpage{385}
(\byear{1997})
\end{barticle}
\endbibitem

\bibitem{rohwer2010kinetic}
\begin{barticle}
\bauthor{\bsnm{Rohwer}, \binits{J.M.}},
\bauthor{\bsnm{Hofmeyr}, \binits{J.-H.S.}}:
\batitle{Kinetic and thermodynamic aspects of enzyme control and regulation}.
\bjtitle{The Journal of Physical Chemistry B}
\bvolume{114}(\bissue{49}),
\bfpage{16280}--\blpage{16289}
(\byear{2010})
\end{barticle}
\endbibitem

\bibitem{sauro2011enzyme}
\begin{bbook}
\bauthor{\bsnm{Sauro}, \binits{H.M.}}:
\bbtitle{Enzyme Kinetics for Systems Biology}.
\bpublisher{Ambrosius Publishing},
\blocation{Seattle}
(\byear{2011})
\end{bbook}
\endbibitem

\bibitem{Ferrell_Ha_2014a}
\begin{barticle}
\bauthor{\bsnm{Ferrell~Jr}, \binits{J.E.}},
\bauthor{\bsnm{Ha}, \binits{S.H.}}:
\batitle{Ultrasensitivity part i: Michaelian responses and zero-order
  ultrasensitivity}.
\bjtitle{Trends in biochemical sciences}
\bvolume{39}(\bissue{10}),
\bfpage{496}--\blpage{503}
(\byear{2014})
\end{barticle}
\endbibitem

\bibitem{Ferrell_Ha_2014b}
\begin{barticle}
\bauthor{\bsnm{Ferrell}, \binits{J.E.}},
\bauthor{\bsnm{Ha}, \binits{S.H.}}, \betal:
\batitle{Ultrasensitivity part ii: multisite phosphorylation, stoichiometric
  inhibitors, and positive feedback}.
\bjtitle{Trends in biochemical sciences}
\bvolume{39}(\bissue{11}),
\bfpage{556}--\blpage{569}
(\byear{2014})
\end{barticle}
\endbibitem

\bibitem{Ferrell_Ha_2014c}
\begin{barticle}
\bauthor{\bsnm{Ferrell}, \binits{J.E.}},
\bauthor{\bsnm{Ha}, \binits{S.H.}}:
\batitle{Ultrasensitivity part iii: cascades, bistable switches, and
  oscillators}.
\bjtitle{Trends in biochemical sciences}
\bvolume{39}(\bissue{12}),
\bfpage{612}--\blpage{618}
(\byear{2014})
\end{barticle}
\endbibitem

\bibitem{straube2017analysis}
\begin{barticle}
\bauthor{\bsnm{Straube}, \binits{R.}}:
\batitle{Analysis of network motifs in cellular regulation: Structural
  similarities, input--output relations and signal integration}.
\bjtitle{Biosystems}
\bvolume{162},
\bfpage{215}--\blpage{232}
(\byear{2017})
\end{barticle}
\endbibitem

\bibitem{Goldbeter_Koshland_1981}
\begin{barticle}
\bauthor{\bsnm{Goldbeter}, \binits{A.}},
\bauthor{\bsnm{Koshland~Jr}, \binits{D.E.}}:
\batitle{An amplified sensitivity arising from covalent modification in
  biological systems.}
\bjtitle{Proceedings of the National Academy of Sciences}
\bvolume{78}(\bissue{11}),
\bfpage{6840}--\blpage{6844}
(\byear{1981})
\end{barticle}
\endbibitem

\bibitem{Bluthgen_Kholodenko_2006}
\begin{barticle}
\bauthor{\bsnm{Bl{\"u}thgen}, \binits{N.}},
\bauthor{\bsnm{Bruggeman}, \binits{F.J.}},
\bauthor{\bsnm{Legewie}, \binits{S.}},
\bauthor{\bsnm{Herzel}, \binits{H.}},
\bauthor{\bsnm{Westerhoff}, \binits{H.V.}},
\bauthor{\bsnm{Kholodenko}, \binits{B.N.}}:
\batitle{Effects of sequestration on signal transduction cascades}.
\bjtitle{The FEBS journal}
\bvolume{273}(\bissue{5}),
\bfpage{895}--\blpage{906}
(\byear{2006})
\end{barticle}
\endbibitem

\bibitem{Hsu_Becskei_2016}
\begin{barticle}
\bauthor{\bsnm{Hsu}, \binits{C.}},
\bauthor{\bsnm{Jaquet}, \binits{V.}},
\bauthor{\bsnm{Gencoglu}, \binits{M.}},
\bauthor{\bsnm{Becskei}, \binits{A.}}:
\batitle{Protein dimerization generates bistability in positive feedback
  loops}.
\bjtitle{Cell reports}
\bvolume{16}(\bissue{5}),
\bfpage{1204}--\blpage{1210}
(\byear{2016})
\end{barticle}
\endbibitem

\bibitem{buchler2009protein}
\begin{barticle}
\bauthor{\bsnm{Buchler}, \binits{N.E.}},
\bauthor{\bsnm{Cross}, \binits{F.R.}}:
\batitle{Protein sequestration generates a flexible ultrasensitive response in
  a genetic network}.
\bjtitle{Molecular systems biology}
\bvolume{5}(\bissue{1}),
\bfpage{272}
(\byear{2009})
\end{barticle}
\endbibitem

\bibitem{Arkin_McAdams_1998}
\begin{barticle}
\bauthor{\bsnm{Arkin}, \binits{A.}},
\bauthor{\bsnm{Ross}, \binits{J.}},
\bauthor{\bsnm{McAdams}, \binits{H.H.}}:
\batitle{Stochastic kinetic analysis of developmental pathway bifurcation in
  phage $\lambda$-infected escherichia coli cells}.
\bjtitle{Genetics}
\bvolume{149}(\bissue{4}),
\bfpage{1633}--\blpage{1648}
(\byear{1998})
\end{barticle}
\endbibitem

\bibitem{Ferrell_Machleder_1998}
\begin{barticle}
\bauthor{\bsnm{Ferrell~Jr}, \binits{J.E.}},
\bauthor{\bsnm{Machleder}, \binits{E.M.}}:
\batitle{The biochemical basis of an all-or-none cell fate switch in xenopus
  oocytes}.
\bjtitle{Science}
\bvolume{280}(\bissue{5365}),
\bfpage{895}--\blpage{898}
(\byear{1998})
\end{barticle}
\endbibitem

\bibitem{Gardner_Collins_2000}
\begin{barticle}
\bauthor{\bsnm{Gardner}, \binits{T.S.}},
\bauthor{\bsnm{Cantor}, \binits{C.R.}},
\bauthor{\bsnm{Collins}, \binits{J.J.}}:
\batitle{Construction of a genetic toggle switch in escherichia coli}.
\bjtitle{Nature}
\bvolume{403}(\bissue{6767}),
\bfpage{339}--\blpage{342}
(\byear{2000})
\end{barticle}
\endbibitem

\bibitem{Fell_Sauro_1985}
\begin{barticle}
\bauthor{\bsnm{Fell}, \binits{D.A.}},
\bauthor{\bsnm{Sauro}, \binits{H.M.}}:
\batitle{Metabolic control and its analysis: additional relationships between
  elasticities and control coefficients}.
\bjtitle{European Journal of Biochemistry}
\bvolume{148}(\bissue{3}),
\bfpage{555}--\blpage{561}
(\byear{1985})
\end{barticle}
\endbibitem

\bibitem{Cornish-Bowden_Cardenas_2001}
\begin{barticle}
\bauthor{\bsnm{Cornish-Bowden}, \binits{A.}},
\bauthor{\bsnm{C{\'a}rdenas}, \binits{M.L.}}:
\batitle{Information transfer in metabolic pathways: effects of irreversible
  steps in computer models}.
\bjtitle{European Journal of Biochemistry}
\bvolume{268}(\bissue{24}),
\bfpage{6616}--\blpage{6624}
(\byear{2001})
\end{barticle}
\endbibitem

\bibitem{Ventura_Merajver_2008}
\begin{barticle}
\bauthor{\bsnm{Ventura}, \binits{A.C.}},
\bauthor{\bsnm{Sepulchre}, \binits{J.-A.}},
\bauthor{\bsnm{Merajver}, \binits{S.D.}}:
\batitle{A hidden feedback in signaling cascades is revealed}.
\bjtitle{PLoS computational biology}
\bvolume{4}(\bissue{3}),
\bfpage{1000041}
(\byear{2008})
\end{barticle}
\endbibitem

\bibitem{Ventura_Ninfa_2010}
\begin{barticle}
\bauthor{\bsnm{Ventura}, \binits{A.C.}},
\bauthor{\bsnm{Jiang}, \binits{P.}},
\bauthor{\bsnm{Van~Wassenhove}, \binits{L.}},
\bauthor{\bsnm{Del~Vecchio}, \binits{D.}},
\bauthor{\bsnm{Merajver}, \binits{S.D.}},
\bauthor{\bsnm{Ninfa}, \binits{A.J.}}:
\batitle{Signaling properties of a covalent modification cycle are altered by a
  downstream target}.
\bjtitle{Proceedings of the National Academy of Sciences}
\bvolume{107}(\bissue{22}),
\bfpage{10032}--\blpage{10037}
(\byear{2010})
\end{barticle}
\endbibitem

\bibitem{Gottesman_1999}
\begin{barticle}
\bauthor{\bsnm{Gottesman}, \binits{S.}}:
\batitle{Regulation by proteolysis: developmental switches}.
\bjtitle{Current opinion in microbiology}
\bvolume{2}(\bissue{2}),
\bfpage{142}--\blpage{147}
(\byear{1999})
\end{barticle}
\endbibitem

\bibitem{Jenal_Hengge-Aronis_2003}
\begin{barticle}
\bauthor{\bsnm{Jenal}, \binits{U.}},
\bauthor{\bsnm{Hengge-Aronis}, \binits{R.}}:
\batitle{Regulation by proteolysis in bacterial cells}.
\bjtitle{Current opinion in microbiology}
\bvolume{6}(\bissue{2}),
\bfpage{163}--\blpage{172}
(\byear{2003})
\end{barticle}
\endbibitem

\bibitem{Lai_2004}
\begin{botherref}
\oauthor{\bsnm{Lai}, \binits{E.C.}}:
Notch signaling: control of cell communication and cell fate
(2004)
\end{botherref}
\endbibitem

\bibitem{Guimera_Amaral_2005}
\begin{barticle}
\bauthor{\bsnm{Guimera}, \binits{R.}},
\bauthor{\bsnm{Nunes~Amaral}, \binits{L.A.}}:
\batitle{Functional cartography of complex metabolic networks}.
\bjtitle{nature}
\bvolume{433}(\bissue{7028}),
\bfpage{895}--\blpage{900}
(\byear{2005})
\end{barticle}
\endbibitem

\bibitem{Clune_Lipson_2013}
\begin{barticle}
\bauthor{\bsnm{Clune}, \binits{J.}},
\bauthor{\bsnm{Mouret}, \binits{J.-B.}},
\bauthor{\bsnm{Lipson}, \binits{H.}}:
\batitle{The evolutionary origins of modularity}.
\bjtitle{Proceedings of the Royal Society b: Biological sciences}
\bvolume{280}(\bissue{1755}),
\bfpage{20122863}
(\byear{2013})
\end{barticle}
\endbibitem

\bibitem{Preissler_Deuerling_2012}
\begin{barticle}
\bauthor{\bsnm{Preissler}, \binits{S.}},
\bauthor{\bsnm{Deuerling}, \binits{E.}}:
\batitle{Ribosome-associated chaperones as key players in proteostasis}.
\bjtitle{Trends in biochemical sciences}
\bvolume{37}(\bissue{7}),
\bfpage{274}--\blpage{283}
(\byear{2012})
\end{barticle}
\endbibitem

\bibitem{Pareek_Benkovic_2021}
\begin{barticle}
\bauthor{\bsnm{Pareek}, \binits{V.}},
\bauthor{\bsnm{Sha}, \binits{Z.}},
\bauthor{\bsnm{He}, \binits{J.}},
\bauthor{\bsnm{Wingreen}, \binits{N.S.}},
\bauthor{\bsnm{Benkovic}, \binits{S.J.}}:
\batitle{Metabolic channeling: predictions, deductions, and evidence}.
\bjtitle{Molecular cell}
\bvolume{81}(\bissue{18}),
\bfpage{3775}--\blpage{3785}
(\byear{2021})
\end{barticle}
\endbibitem

\bibitem{Langeberg_Scott_2015}
\begin{barticle}
\bauthor{\bsnm{Langeberg}, \binits{L.K.}},
\bauthor{\bsnm{Scott}, \binits{J.D.}}:
\batitle{Signalling scaffolds and local organization of cellular behaviour}.
\bjtitle{Nature reviews Molecular cell biology}
\bvolume{16}(\bissue{4}),
\bfpage{232}--\blpage{244}
(\byear{2015})
\end{barticle}
\endbibitem

\bibitem{Csete_Doyle_2002}
\begin{barticle}
\bauthor{\bsnm{Csete}, \binits{M.E.}},
\bauthor{\bsnm{Doyle}, \binits{J.C.}}:
\batitle{Reverse engineering of biological complexity}.
\bjtitle{science}
\bvolume{295}(\bissue{5560}),
\bfpage{1664}--\blpage{1669}
(\byear{2002})
\end{barticle}
\endbibitem

\bibitem{Franco_Murray_2009}
\begin{bchapter}
\bauthor{\bsnm{Franco}, \binits{E.}},
\bauthor{\bsnm{Del~Vecchio}, \binits{D.}},
\bauthor{\bsnm{Murray}, \binits{R.M.}}:
\bctitle{Design of insulating devices for in vitro synthetic circuits}.
In: \bbtitle{Proceedings of the 48h IEEE Conference on Decision and Control
  (CDC) Held Jointly with 2009 28th Chinese Control Conference},
pp. \bfpage{4584}--\blpage{4589}
(\byear{2009}).
\bcomment{IEEE}
\end{bchapter}
\endbibitem

\bibitem{Shah_DelVecchio_2016}
\begin{bchapter}
\bauthor{\bsnm{Shah}, \binits{R.}},
\bauthor{\bsnm{Del~Vecchio}, \binits{D.}}:
\bctitle{An n-stage cascade of phosphorylation cycles as an insulation device
  for synthetic biological circuits}.
In: \bbtitle{2016 European Control Conference (ECC)},
pp. \bfpage{1832}--\blpage{1837}
(\byear{2016}).
\bcomment{IEEE}
\end{bchapter}
\endbibitem

\bibitem{Ferrell_2009}
\begin{barticle}
\bauthor{\bsnm{Ferrell}, \binits{J.E.}}:
\batitle{Signaling motifs and weber's law}.
\bjtitle{Molecular cell}
\bvolume{36}(\bissue{5}),
\bfpage{724}--\blpage{727}
(\byear{2009})
\end{barticle}
\endbibitem

\bibitem{Bar-Even_Barkai_2006}
\begin{barticle}
\bauthor{\bsnm{Bar-Even}, \binits{A.}},
\bauthor{\bsnm{Paulsson}, \binits{J.}},
\bauthor{\bsnm{Maheshri}, \binits{N.}},
\bauthor{\bsnm{Carmi}, \binits{M.}},
\bauthor{\bsnm{O'Shea}, \binits{E.}},
\bauthor{\bsnm{Pilpel}, \binits{Y.}},
\bauthor{\bsnm{Barkai}, \binits{N.}}:
\batitle{Noise in protein expression scales with natural protein abundance}.
\bjtitle{Nature genetics}
\bvolume{38}(\bissue{6}),
\bfpage{636}--\blpage{643}
(\byear{2006})
\end{barticle}
\endbibitem

\bibitem{Shoval_Alon_2010}
\begin{barticle}
\bauthor{\bsnm{Shoval}, \binits{O.}},
\bauthor{\bsnm{Goentoro}, \binits{L.}},
\bauthor{\bsnm{Hart}, \binits{Y.}},
\bauthor{\bsnm{Mayo}, \binits{A.}},
\bauthor{\bsnm{Sontag}, \binits{E.}},
\bauthor{\bsnm{Alon}, \binits{U.}}:
\batitle{Fold-change detection and scalar symmetry of sensory input fields}.
\bjtitle{Proceedings of the National Academy of Sciences}
\bvolume{107}(\bissue{36}),
\bfpage{15995}--\blpage{16000}
(\byear{2010})
\end{barticle}
\endbibitem

\bibitem{Webre_Stock_2003}
\begin{barticle}
\bauthor{\bsnm{Webre}, \binits{D.J.}},
\bauthor{\bsnm{Wolanin}, \binits{P.M.}},
\bauthor{\bsnm{Stock}, \binits{J.B.}}:
\batitle{Bacterial chemotaxis}.
\bjtitle{Current Biology}
\bvolume{13}(\bissue{2}),
\bfpage{47}--\blpage{49}
(\byear{2003})
\end{barticle}
\endbibitem

\bibitem{Tu_Berg_2008}
\begin{barticle}
\bauthor{\bsnm{Tu}, \binits{Y.}},
\bauthor{\bsnm{Shimizu}, \binits{T.S.}},
\bauthor{\bsnm{Berg}, \binits{H.C.}}:
\batitle{Modeling the chemotactic response of escherichia coli to time-varying
  stimuli}.
\bjtitle{Proceedings of the National Academy of Sciences}
\bvolume{105}(\bissue{39}),
\bfpage{14855}--\blpage{14860}
(\byear{2008})
\end{barticle}
\endbibitem

\bibitem{Clausznitzer_Endres_2014}
\begin{barticle}
\bauthor{\bsnm{Clausznitzer}, \binits{D.}},
\bauthor{\bsnm{Micali}, \binits{G.}},
\bauthor{\bsnm{Neumann}, \binits{S.}},
\bauthor{\bsnm{Sourjik}, \binits{V.}},
\bauthor{\bsnm{Endres}, \binits{R.G.}}:
\batitle{Predicting chemical environments of bacteria from receptor signaling}.
\bjtitle{PLoS Computational Biology}
\bvolume{10}(\bissue{10}),
\bfpage{1003870}
(\byear{2014})
\end{barticle}
\endbibitem

\bibitem{Lee_Gaudet_2014}
\begin{barticle}
\bauthor{\bsnm{Lee}, \binits{R.E.}},
\bauthor{\bsnm{Walker}, \binits{S.R.}},
\bauthor{\bsnm{Savery}, \binits{K.}},
\bauthor{\bsnm{Frank}, \binits{D.A.}},
\bauthor{\bsnm{Gaudet}, \binits{S.}}:
\batitle{Fold change of nuclear nf-$\kappa$b determines tnf-induced
  transcription in single cells}.
\bjtitle{Molecular cell}
\bvolume{53}(\bissue{6}),
\bfpage{867}--\blpage{879}
(\byear{2014})
\end{barticle}
\endbibitem

\bibitem{Cohen-Saidon_Alon_2009}
\begin{barticle}
\bauthor{\bsnm{Cohen-Saidon}, \binits{C.}},
\bauthor{\bsnm{Cohen}, \binits{A.A.}},
\bauthor{\bsnm{Sigal}, \binits{A.}},
\bauthor{\bsnm{Liron}, \binits{Y.}},
\bauthor{\bsnm{Alon}, \binits{U.}}:
\batitle{Dynamics and variability of erk2 response to egf in individual living
  cells}.
\bjtitle{Molecular cell}
\bvolume{36}(\bissue{5}),
\bfpage{885}--\blpage{893}
(\byear{2009})
\end{barticle}
\endbibitem

\bibitem{Goentoro_Kirschner_2009}
\begin{barticle}
\bauthor{\bsnm{Goentoro}, \binits{L.}},
\bauthor{\bsnm{Kirschner}, \binits{M.W.}}:
\batitle{Evidence that fold-change, and not absolute level, of $\beta$-catenin
  dictates wnt signaling}.
\bjtitle{Molecular cell}
\bvolume{36}(\bissue{5}),
\bfpage{872}--\blpage{884}
(\byear{2009})
\end{barticle}
\endbibitem

\bibitem{Hart_Alon_2013b}
\begin{barticle}
\bauthor{\bsnm{Hart}, \binits{Y.}},
\bauthor{\bsnm{Mayo}, \binits{A.E.}},
\bauthor{\bsnm{Shoval}, \binits{O.}},
\bauthor{\bsnm{Alon}, \binits{U.}}:
\batitle{Comparing apples and oranges: fold-change detection of multiple
  simultaneous inputs}.
\bjtitle{PloS one}
\bvolume{8}(\bissue{3}),
\bfpage{57455}
(\byear{2013})
\end{barticle}
\endbibitem

\bibitem{Shinar_Alon_2007}
\begin{barticle}
\bauthor{\bsnm{Shinar}, \binits{G.}},
\bauthor{\bsnm{Milo}, \binits{R.}},
\bauthor{\bsnm{Mart{\'\i}nez}, \binits{M.R.}},
\bauthor{\bsnm{Alon}, \binits{U.}}:
\batitle{Input--output robustness in simple bacterial signaling systems}.
\bjtitle{Proceedings of the National Academy of Sciences}
\bvolume{104}(\bissue{50}),
\bfpage{19931}--\blpage{19935}
(\byear{2007})
\end{barticle}
\endbibitem

\bibitem{Hart_Alon_2013a}
\begin{barticle}
\bauthor{\bsnm{Hart}, \binits{Y.}},
\bauthor{\bsnm{Alon}, \binits{U.}}:
\batitle{The utility of paradoxical components in biological circuits}.
\bjtitle{Molecular cell}
\bvolume{49}(\bissue{2}),
\bfpage{213}--\blpage{221}
(\byear{2013})
\end{barticle}
\endbibitem

\bibitem{Henderson_Lew_2019}
\begin{barticle}
\bauthor{\bsnm{Henderson}, \binits{N.T.}},
\bauthor{\bsnm{Pablo}, \binits{M.}},
\bauthor{\bsnm{Ghose}, \binits{D.}},
\bauthor{\bsnm{Clark-Cotton}, \binits{M.R.}},
\bauthor{\bsnm{Zyla}, \binits{T.R.}},
\bauthor{\bsnm{Nolen}, \binits{J.}},
\bauthor{\bsnm{Elston}, \binits{T.C.}},
\bauthor{\bsnm{Lew}, \binits{D.J.}}:
\batitle{Ratiometric gpcr signaling enables directional sensing in yeast}.
\bjtitle{PLoS biology}
\bvolume{17}(\bissue{10}),
\bfpage{3000484}
(\byear{2019})
\end{barticle}
\endbibitem

\bibitem{Russo_Silhavy_1993}
\begin{barticle}
\bauthor{\bsnm{Russo}, \binits{F.D.}},
\bauthor{\bsnm{Silhavy}, \binits{T.J.}}:
\batitle{The essential tension: opposed reactions in bacterial two-component
  regulatory systems}.
\bjtitle{Trends in microbiology}
\bvolume{1}(\bissue{8}),
\bfpage{306}--\blpage{310}
(\byear{1993})
\end{barticle}
\endbibitem

\bibitem{Batchelor_Goulian_2003}
\begin{barticle}
\bauthor{\bsnm{Batchelor}, \binits{E.}},
\bauthor{\bsnm{Goulian}, \binits{M.}}:
\batitle{Robustness and the cycle of phosphorylation and dephosphorylation in a
  two-component regulatory system}.
\bjtitle{Proceedings of the National Academy of Sciences}
\bvolume{100}(\bissue{2}),
\bfpage{691}--\blpage{696}
(\byear{2003})
\end{barticle}
\endbibitem

\bibitem{Song_Kumar_2012}
\begin{barticle}
\bauthor{\bsnm{Song}, \binits{Q.}},
\bauthor{\bsnm{Kumar}, \binits{A.}}:
\batitle{An overview of autophagy and yeast pseudohyphal growth: integration of
  signaling pathways during nitrogen stress}.
\bjtitle{Cells}
\bvolume{1}(\bissue{3}),
\bfpage{263}--\blpage{283}
(\byear{2012})
\end{barticle}
\endbibitem

\bibitem{Chen_Kurjan_1997}
\begin{barticle}
\bauthor{\bsnm{Chen}, \binits{T.}},
\bauthor{\bsnm{Kurjan}, \binits{J.}}:
\batitle{Saccharomyces cerevisiae mpt5p interacts with sst2p and plays roles in
  pheromone sensitivity and recovery from pheromone arrest}.
\bjtitle{Molecular and Cellular Biology}
\bvolume{17}(\bissue{6}),
\bfpage{3429}--\blpage{3439}
(\byear{1997})
\end{barticle}
\endbibitem

\bibitem{Poritz_Kamb_2001}
\begin{barticle}
\bauthor{\bsnm{Poritz}, \binits{M.A.}},
\bauthor{\bsnm{Malmstrom}, \binits{S.}},
\bauthor{\bsnm{Kim}, \binits{M.K.-H.}},
\bauthor{\bsnm{Rossmeissl}, \binits{P.J.}},
\bauthor{\bsnm{Kamb}, \binits{A.}}:
\batitle{Graded mode of transcriptional induction in yeast pheromone signalling
  revealed by single-cell analysis}.
\bjtitle{Yeast}
\bvolume{18}(\bissue{14}),
\bfpage{1331}--\blpage{1338}
(\byear{2001})
\end{barticle}
\endbibitem

\bibitem{Jenness_Hartwell_1986}
\begin{barticle}
\bauthor{\bsnm{Jenness}, \binits{D.D.}},
\bauthor{\bsnm{Burkholder}, \binits{A.C.}},
\bauthor{\bsnm{Hartwell}, \binits{L.H.}}:
\batitle{Binding of $\alpha$-factor pheromone to saccharomyces cerevisiae a
  cells: dissociation constant and number of binding sites}.
\bjtitle{Molecular and cellular biology}
\bvolume{6}(\bissue{1}),
\bfpage{318}--\blpage{320}
(\byear{1986})
\end{barticle}
\endbibitem

\bibitem{Bajaj_Dumont_2004}
\begin{barticle}
\bauthor{\bsnm{Bajaj}, \binits{A.}},
\bauthor{\bsnm{{\'C}eli{\'c}}, \binits{A.}},
\bauthor{\bsnm{Ding}, \binits{F.-X.}},
\bauthor{\bsnm{Naider}, \binits{F.}},
\bauthor{\bsnm{Becker}, \binits{J.M.}},
\bauthor{\bsnm{Dumont}, \binits{M.E.}}:
\batitle{A fluorescent $\alpha$-factor analogue exhibits multiple steps on
  binding to its g protein coupled receptor in yeast}.
\bjtitle{Biochemistry}
\bvolume{43}(\bissue{42}),
\bfpage{13564}--\blpage{13578}
(\byear{2004})
\end{barticle}
\endbibitem

\bibitem{Kofahl_Klipp_2004}
\begin{barticle}
\bauthor{\bsnm{Kofahl}, \binits{B.}},
\bauthor{\bsnm{Klipp}, \binits{E.}}:
\batitle{Modelling the dynamics of the yeast pheromone pathway}.
\bjtitle{Yeast}
\bvolume{21}(\bissue{10}),
\bfpage{831}--\blpage{850}
(\byear{2004})
\end{barticle}
\endbibitem

\bibitem{Suderman_Deeds_2013}
\begin{barticle}
\bauthor{\bsnm{Suderman}, \binits{R.}},
\bauthor{\bsnm{Deeds}, \binits{E.J.}}:
\batitle{Machines vs. ensembles: effective mapk signaling through heterogeneous
  sets of protein complexes}.
\bjtitle{PLoS computational biology}
\bvolume{9}(\bissue{10}),
\bfpage{1003278}
(\byear{2013})
\end{barticle}
\endbibitem

\bibitem{Olson_Sadowski_2000}
\begin{barticle}
\bauthor{\bsnm{Olson}, \binits{K.A.}},
\bauthor{\bsnm{Nelson}, \binits{C.}},
\bauthor{\bsnm{Tai}, \binits{G.}},
\bauthor{\bsnm{Hung}, \binits{W.}},
\bauthor{\bsnm{Yong}, \binits{C.}},
\bauthor{\bsnm{Astell}, \binits{C.}},
\bauthor{\bsnm{Sadowski}, \binits{I.}}:
\batitle{Two regulators of ste12p inhibit pheromone-responsive transcription by
  separate mechanisms}.
\bjtitle{Molecular and cellular biology}
\bvolume{20}(\bissue{12}),
\bfpage{4199}--\blpage{4209}
(\byear{2000})
\end{barticle}
\endbibitem

\bibitem{Cook_Thorner_1997}
\begin{barticle}
\bauthor{\bsnm{Cook}, \binits{J.G.}},
\bauthor{\bsnm{Bardwell}, \binits{L.}},
\bauthor{\bsnm{Thorner}, \binits{J.}}:
\batitle{Inhibitory and activating functions for mapk kss1 in the s. cerevisiae
  filamentous-growth signalling pathway}.
\bjtitle{Nature}
\bvolume{390}(\bissue{6655}),
\bfpage{85}--\blpage{88}
(\byear{1997})
\end{barticle}
\endbibitem

\bibitem{Hao_Dohlman_2008}
\begin{barticle}
\bauthor{\bsnm{Hao}, \binits{N.}},
\bauthor{\bsnm{Nayak}, \binits{S.}},
\bauthor{\bsnm{Behar}, \binits{M.}},
\bauthor{\bsnm{Shanks}, \binits{R.H.}},
\bauthor{\bsnm{Nagiec}, \binits{M.J.}},
\bauthor{\bsnm{Errede}, \binits{B.}},
\bauthor{\bsnm{Hasty}, \binits{J.}},
\bauthor{\bsnm{Elston}, \binits{T.C.}},
\bauthor{\bsnm{Dohlman}, \binits{H.G.}}:
\batitle{Regulation of cell signaling dynamics by the protein kinase-scaffold
  ste5}.
\bjtitle{Molecular cell}
\bvolume{30}(\bissue{5}),
\bfpage{649}--\blpage{656}
(\byear{2008})
\end{barticle}
\endbibitem

\bibitem{sauro2004quantitative}
\begin{barticle}
\bauthor{\bsnm{Sauro}, \binits{H.M.}},
\bauthor{\bsnm{Kholodenko}, \binits{B.N.}}:
\batitle{Quantitative analysis of signaling networks}.
\bjtitle{Progress in biophysics and molecular biology}
\bvolume{86}(\bissue{1}),
\bfpage{5}--\blpage{43}
(\byear{2004})
\end{barticle}
\endbibitem

\bibitem{lim2013design}
\begin{barticle}
\bauthor{\bsnm{Lim}, \binits{W.A.}},
\bauthor{\bsnm{Lee}, \binits{C.M.}},
\bauthor{\bsnm{Tang}, \binits{C.}}:
\batitle{Design principles of regulatory networks: searching for the molecular
  algorithms of the cell}.
\bjtitle{Molecular cell}
\bvolume{49}(\bissue{2}),
\bfpage{202}--\blpage{212}
(\byear{2013})
\end{barticle}
\endbibitem

\bibitem{Gould_1989}
\begin{bbook}
\bauthor{\bsnm{Gould}, \binits{S.J.}}:
\bbtitle{Wonderful Life: {The} Burgess Shale and the Nature of History}.
\bpublisher{WW Norton \& Company},
\blocation{New York}
(\byear{1989})
\end{bbook}
\endbibitem

\bibitem{Tomita_2001}
\begin{barticle}
\bauthor{\bsnm{Tomita}, \binits{M.}}:
\batitle{Whole-cell simulation: a grand challenge of the 21st century}.
\bjtitle{TRENDS in Biotechnology}
\bvolume{19}(\bissue{6}),
\bfpage{205}--\blpage{210}
(\byear{2001})
\end{barticle}
\endbibitem

\bibitem{Karr_Covert_2012}
\begin{barticle}
\bauthor{\bsnm{Karr}, \binits{J.R.}},
\bauthor{\bsnm{Sanghvi}, \binits{J.C.}},
\bauthor{\bsnm{Macklin}, \binits{D.N.}},
\bauthor{\bsnm{Gutschow}, \binits{M.V.}},
\bauthor{\bsnm{Jacobs}, \binits{J.M.}},
\bauthor{\bsnm{Bolival~Jr}, \binits{B.}},
\bauthor{\bsnm{Assad-Garcia}, \binits{N.}},
\bauthor{\bsnm{Glass}, \binits{J.I.}},
\bauthor{\bsnm{Covert}, \binits{M.W.}}:
\batitle{A whole-cell computational model predicts phenotype from genotype}.
\bjtitle{Cell}
\bvolume{150}(\bissue{2}),
\bfpage{389}--\blpage{401}
(\byear{2012})
\end{barticle}
\endbibitem

\end{thebibliography}

\end{document}